\documentclass[aps,prd,twocolumn,showpacs,10pt,superscriptaddress,preprintnumbers,nofootinbib]{revtex4-1}
\pdfoutput=1
\usepackage[english]{babel}
\usepackage{amsmath}
\usepackage{graphicx}
\usepackage[
  left=1.2cm,
  right=1.2cm,
  bottom=1.5cm,
  bottom=1.5cm
]{geometry}
\usepackage{bm}
\usepackage{booktabs}
\usepackage{slashed}
\usepackage{multirow}
\usepackage{booktabs}
\usepackage{xspace}
\usepackage{hyperref}
\usepackage[usenames,dvipsnames]{xcolor}
\usepackage{feynmp-auto}
\usepackage[ruled,longend]{algorithm2e}

\allowdisplaybreaks[4]

\newcommand{\beq}{\begin{equation}\begin{aligned}}
\newcommand{\eeq}{\end{aligned}\end{equation}}

\definecolor{darkyellow}{rgb}{0.5, 0.5, 0.0}
\definecolor{darkpurple}{rgb}{0.5, 0.2, 0.8}
\definecolor{darkblue}{rgb}{0.0, 0.0, 0.8}
\definecolor{darkgreen}{rgb}{0.0, 0.4, 0.0}
\definecolor{darkred}{rgb}{0.5, 0.0, 0.0}

\hypersetup{
    linktocpage,
     colorlinks,
     citecolor=darkgreen,
     linkcolor= darkgreen,
     urlcolor=darkgreen
}

\begin{document}
\title{Pushing the Limits of Pulse Shape Discrimination in a Large Liquid Xenon Detector}



\author{D.S.~Akerib}
\affiliation{SLAC National Accelerator Laboratory, Menlo Park, CA 94025-7015, USA}
\affiliation{Kavli Institute for Particle Astrophysics and Cosmology, Stanford University, Stanford, CA  94305-4085 USA}

\author{A.K.~Al Musalhi}
\affiliation{University College London (UCL), Department of Physics and Astronomy, London WC1E 6BT, UK}

\author{F.~Alder}
\affiliation{University College London (UCL), Department of Physics and Astronomy, London WC1E 6BT, UK}

\author{B.J.~Almquist}
\affiliation{Brown University, Department of Physics, Providence, RI 02912-9037, USA}

\author{C.S.~Amarasinghe}
\affiliation{University of California, Santa Barbara, Department of Physics, Santa Barbara, CA 93106-9530, USA}

\author{A.~Ames}
\affiliation{SLAC National Accelerator Laboratory, Menlo Park, CA 94025-7015, USA}
\affiliation{Kavli Institute for Particle Astrophysics and Cosmology, Stanford University, Stanford, CA  94305-4085 USA}

\author{T.J.~Anderson}
\affiliation{SLAC National Accelerator Laboratory, Menlo Park, CA 94025-7015, USA}
\affiliation{Kavli Institute for Particle Astrophysics and Cosmology, Stanford University, Stanford, CA  94305-4085 USA}

\author{N.~Angelides}
\affiliation{University of Zurich, Department of Physics, 8057 Zurich, Switzerland}

\author{H.M.~Ara\'{u}jo}
\affiliation{Imperial College London, Physics Department, Blackett Laboratory, London SW7 2AZ, UK}
\affiliation{STFC Rutherford Appleton Laboratory (RAL), Didcot, OX11 0QX, UK}

\author{J.E.~Armstrong}
\affiliation{University of Maryland, Department of Physics, College Park, MD 20742-4111, USA}

% ...

\collaboration{The LZ Collaboration}

\begin{abstract}
    The LUX-ZEPLIN (LZ) experiment is a direct-detection dark matter experiment, optimized to search for weakly interacting massive particles (WIMPs) through WIMP-nucleon interactions. The main challenge in dark matter detection is differentiating between WIMP signals and background events. In LZ, the ratio of ionization to scintillation signals (charge-to-light) is the primary method for rejecting electronic recoil (ER) background. Pulse shape discrimination (PSD) offers a method for additional ER backgrounds rejection in liquid xenon detectors. In this paper, the discrimination power of PSD with the LZ experiment is discussed. To precisely characterize the scintillation pulse shape, an analysis framework is developed to reconstruct the detection time of individual photons. Using LZ calibration data, the photon-timing tail fraction discriminator is optimized and achieves ER leakage as low as $15\%$. For specific background processes such as $^{124}$Xe double electron capture, the leakage is reduced further to about $5\%$.
    PSD is combined with charge-to-light to form two-factor discrimination (TFD). The optimized TFD performance is compared with the performance of the charge-to-light method, with the corresponding false positive rate reduced by up to a factor of two for large scintillation pulses.
    Finally, PSD and TFD are applied to data from LZ's WS2024 run and their performance is summarized. 
\end{abstract}

\maketitle
\phantom{x}

\section{Introduction}

There is observational evidence for the existence of dark matter at 
different distance scales from rotation curves~\cite{Rubin:1978kmz} at the galactic scale to the isotropy of the cosmic microwave background~\cite{Holtzman:1989ki} at the cosmological scale.
Weakly Interacting Massive Particles (WIMPs) are 
a well-motivated candidate for dark matter.
Various direct-detection dark matter experiments are searching for WIMPs by detecting the interactions of WIMPs with nuclei. Due to their sensitivity to both spin-independent and spin-dependent WIMP interactions, liquid noble elements are used as target materials, especially liquid xenon (LXe)~\cite{Mount:2017qzi}.

The LUX-ZEPLIN (LZ) experiment is a direct-detection dark matter experiment, located 4850~ft below the surface (4300 mwe) at the Sanford Underground Research Facility (SURF) in Lead, South Dakota. The center of LZ is a dual-phase Time Projection Chamber (TPC) which is surrounded by two veto detectors: the Skin detector and the Outer Detector (OD). The entire LZ detector is shielded by 238~tonnes of ultrapure water~\cite{LZ:2019sgr}. 

The TPC is a cylinder with a diameter and height of 1.5~m, containing 7~tonnes of LXe in its active volume.
The top and bottom of the TPC are instrumented with 494 3-inch Hamamatsu R11410-22 photomultiplier tubes (PMTs), distributed over two arrays. These PMTs detect signals produced by Xe recoils and background signals. 
Interactions
in the LXe produce scintillation photons and ionization electrons. The scintillation light is detected first and produces a prompt scintillation signal (S1). The secondary electroluminescence signal (S2) is created by the ionization electrons that drift to the liquid surface and are extracted into the gas phase. Both S1 and S2 pulses are distributed over multiple PMTs in the top and bottom arrays. The event's $(x,y)$ position is reconstructed using the spatial distribution of the S2 signal on the top array. The time difference between the S1 and S2 pulses provides information about the $z$-position of the interaction. Multiple radioactive sources are used to calibrate the detector response, including tritiated methane (CH$_3$T), $^{83m}$Kr, $^{131m}$Xe, deuterium-deuterium (DD) neutrons, and americium-beryllium (AmBe) neutrons~\cite{LZ:calibration}. LZ began taking data in December 2021. This analysis focuses on the WIMP search (WS) datasets collected between March 2023 and April 2024, referred to as WS2024. Events which pass all data quality (DQ) cuts are referred to as WS events.
More details of the WS2024 dataset and the DQ cuts applied can be found in Ref.~\cite{LZ-WS2024Results}.

The main challenge of dark matter detection is the discrimination between WIMP signals and background events. In LZ, the underground environment and the water shield reduce the backgrounds from cosmic rays and environmental sources.  
The Skin detector and the OD can provide veto signals to further reject background events. The most prominent remaining backgrounds are mainly from gamma and beta radiation released by the residual radioactivity in the detector materials. Since these background events produce primarily electronic recoil (ER) events, discrimination between nuclear recoil (NR) and ER events is key to the observation of WIMPs. 
ER events predominately arise from interactions with the atomic electrons of xenon, while NR events result from interactions with the xenon nuclei.
For most LXe experiments including LZ, the primary discrimination method is charge-to-light discrimination, which achieves a high rejection efficiency for ER events.
To characterize and validate NR and ER discrimination within the WS region of interest (ROI), CH$_3$T and DD calibrations are used.

Despite the strong discrimination power of charge-to-light discrimination, further improvements are desirable to enhance the signal acceptance and achieve stronger sensitivity limits. A second and independent ER background rejection technique, known as Pulse Shape Discrimination (PSD), provides additional rejection capabilities. PSD differentiates NR events from ER events based on the difference in the 
S1 shape due to each preferentially exciting different states.
PSD is the main method to reject ER backgrounds in liquid argon (LAr) experiments~\cite{PhysRevC.78.035801, SCENE:2014iyj,Manthos:2023swh}.
While PSD is not used as a standalone method in LZ, it may be integrated with charge-to-light discrimination as an extra dimension.

In this paper, the discrimination power of PSD in LZ is discussed. Section~\ref{Sec: Theory} introduces the theoretical basis of PSD. Section~\ref{Sec: Corrections} describes how to obtain accurate photon timing spectra for S1 pulses. Section~\ref{Sec: PSD} discusses the optimization of the pulse shape discriminator based on differences in pulse shapes between ER and NR calibration events, along with an evaluation of the performance of PSD. 
This section also introduces the $z$-correction used to account for variations in photon transit times arising from different event locations, enabling PSD to be applied across the full detector volume. 
In Sec.~\ref{Sec: NEST}, the construction and tuning of the photon timing model in NEST is discussed, and the performance of PSD with NEST simulated events is studied. 
Section~\ref{Sec: TFD} introduces the combination of PSD with charge-to-light discrimination to form the two-factor discriminator (TFD), detailing its optimization and performance. Finally, in Sec.~\ref{Sec: WS}, PSD and TFD are applied to WS2024 data and their performance is described. The paper is summarized in Sec.~\ref{Sec: conclusion}.

\section{Theory} \label{Sec: Theory}

For noble elements, scintillation light can be produced through either direct excitation (Eq.\eqref{Eq:excitation}), or recombination of electron-ion pairs (Eq.\eqref{Eq:recombination})~\cite{FUJII2015293}.
\begin{align}
\label{Eq:excitation}
    \text{Xe}^* + \text{Xe} &\rightarrow \text{Xe}^*_2 \notag \\
     \text{Xe}^*_2 &\rightarrow \text{Xe} + \text{Xe} + \gamma
\end{align}
\begin{align}
\label{Eq:recombination}
    \text{Xe}^+ + \text{Xe} &\rightarrow \text{Xe}^+_2 \notag \\
     \text{Xe}^+_2 + e^- &\rightarrow \text{Xe}^{**} + \text{Xe} \notag \\
     \text{Xe}^{**} &\rightarrow \text{Xe}^* + \text{heat}  \\
     \text{Xe}^* + \text{Xe} &\rightarrow \text{Xe}^*_2 \notag \\
     \text{Xe}^*_2 &\rightarrow \text{Xe} + \text{Xe} + \gamma   \notag 
\end{align}
The excited dimer $\text{Xe}^*_2$, producing the scintillation light, can exist in either a singlet or a triplet molecular state. These states have different lifetimes, represented by $\tau _1$ and $\tau _3$. The ratio of singlet to triplet emission is different for NR and ER events, with NR events exhibiting a higher ratio. This difference can be exploited to discriminate NR events from ER events.

In LAr, $\tau _1$ and $\tau _3$ are measured to be $7.0 \pm 1.0 $~ns and $1600 \pm 100 $~ns, respectively~\cite{PhysRevC.78.035801}. This large difference allows the use of PSD as the dominant method of ER background rejection in LAr dark matter detectors. In contrast, previous studies in LXe~\cite{Akerib_2018, PhysRevB.27.5279, PhysRevB.20.3486} report $\tau _1$ values of $2-4 $~ns and $\tau _3$ values of $21-28 $~ns. The small difference between $\tau _1$ and $\tau _3$ in LXe makes PSD more challenging for LXe detectors. Despite this limitation, PSD has been previously studied in various LXe dark matter experiments~\cite{AKIMOV2002245, Kwong_2010, Namwongsa_2017, Hogenbirk_2018,Akerib_2018, Akerib_2020}.

\section{Photon Timing} \label{Sec: Corrections}

The performance of PSD is constrained by several factors: the precision of relative timing offsets among the TPC PMTs, the accuracy in determining the arrival times of individual photons, and the position-dependent variation in photon transit times. 
The sampling rate of the 14-bit 2V ADC used in LZ data acquisition (DAQ) system is 100 MHz and a typical single photoelectron (SPHE) pulse has an amplitude of $49\pm3$ ADC~Counts (ADCC), where 1 ADCC is equivalent to 0.122 mV in LZ DAQ~\cite{LZ:2024bvw}, and a full width at tenth maximum (FWTM) of 60 ns. This section discusses the measurement of relative timing offsets between PMTs and introduces the $N$-photon model used to determine the arrival times of individual photons.

\subsection{Relative Timing} \label{Sec: Relative timing}
Differences in operating voltages, signal cable lengths, and propagation times within the electronics, result in timing differences of the signals from different PMTs. To correct for these differences, channel-to-channel timing calibrations were carried out using the LED calibration system~\cite{LZ:calibration}. For each measurement, only one LED is strobed at a time in order to minimize variations in the flight path between the LED and individual PMTs. 
The LEDs are embedded among the PMTs in the two arrays~\cite{LZ:calibration}. As a result, the LEDs in the top array do not illuminate all PMTs of the top array, and can only be used to measure the timing differences among the bottom PMTs. For the same reason, the bottom LEDs can only be used to determine the timing differences among the top PMTs.


To determine the timing differences, each SPHE waveform from an LED calibration is interpolated, and for each PMT, all interpolation functions are averaged to construct SPHE templates. This template is then fitted to each waveform of the corresponding PMT to determine its arrival time. For each PMT, a photon arrival time distribution is generated, and the time of the $20\%$ rising edge of this distribution is used as its reference time. 
The arrival time distributions are corrected for the time-of-flight (TOF) between the LED used and the PMT. The final timing corrections are determined by calculating the difference between the reference time of each PMT and the reference time of the earliest PMT in the measurement. The resulting channel-to-channel timing corrections have a accuracy of better than 1~ns.

\subsection{$N$-photon Model} \label{Sec: N-photon model}
Waveforms in individual PMTs can contain multiple overlapping photons. To accurately determine the number of photons and their individual arrival times, the $N$-photon model was developed, where `$N$' represents the number of detected photons in an individual PMT. This model reconstructs waveforms by adding $N$ SPHE templates with different amplitudes and time offsets:

\begin{equation}
\label{eq: n-ph}
  \phi (t) = \sum_{i=1}^{N} a_i\times f(t-t_i)+c
\end{equation}
where $f(t)$ is the SPHE template, $a_i$ represents the amplitude of photon $i$, $t_i$ is the time of arrival of photon $i$, and $c$ is an offset introduced to account for baseline noise. The SPHE template for each PMT is obtained from the LED calibrations discussed in Sec.~\ref{Sec: Relative timing}. The parameters $a_i$, $t_i$, and $c$ are determined by fitting Eq.\eqref{eq: n-ph} to each waveform. The parameter $c$ is expected to be 0 and is constrained to a value between $-5$ and $+5$~ADCC.
The arrival time of photon $i$ is taken to be $t_i$, and its pulse area is calculated based on the values of $a_i$ and template areas.

\begin{figure}[h]
    \centering
    \includegraphics[width=0.5\textwidth]{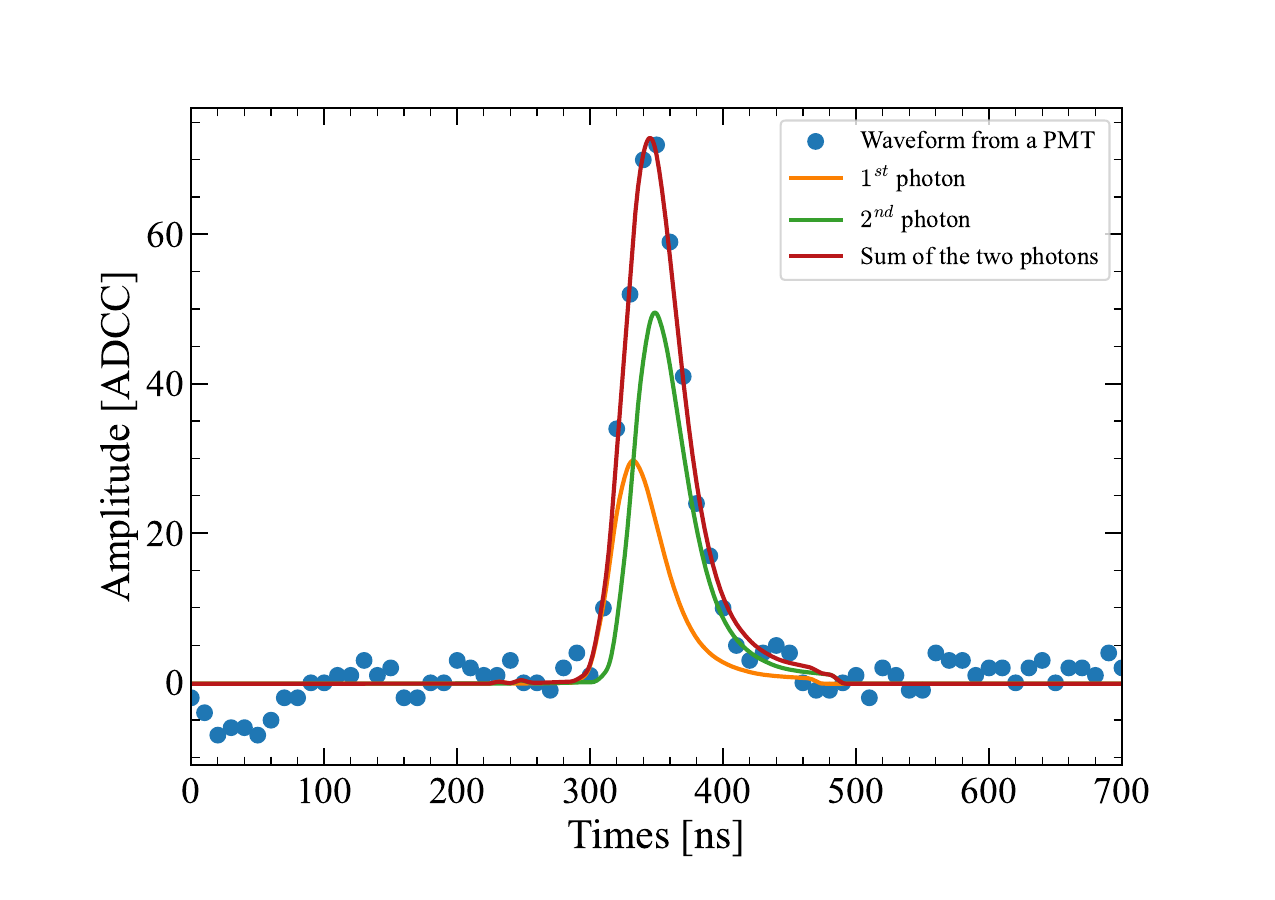}
    \caption{An example of a waveform (blue, dotted) from a WS2024 event processed with a 2-photon model. The result of the fit is shown by the two photons (green and orange). The difference in pulse area between the two photons is consistent with the $\sim30\%$ PMT resolution and the possibility of double-photoelectron emission in the PMT.
    The sum of the two photon distributions is also shown (red). The arrival time of each photon is represented by the peak time.}
    \label{Fig: N-photon Model Example}
\end{figure}

To determine the maximum value of $N$, the area distribution of waveforms from WIMP search events is analyzed. The fraction of PMT waveforms with an area smaller than 3~phd is 98.9$\%$, leading to our choice of $N_{max}=3$ for all waveforms. For each PMT waveform, the 1-, 2-, and 3-photon models are evaluated.
The best value of $N$, given the observed data, is determined using Bayes Theorem.
The likelihood of each model is the product of the maximum likelihood from the fit results and the prior probability derived from the PMT waveform. Given a waveform ($D$) and an $N$-photon model ($M_N$), the probability of the $N$-photon model given the waveform, $P(M_N|D)$ \cite{Akerib_2018}, is given by: 
\begin{equation}
    P(M_N|D)=\frac{P(D|M_N)P(M_N)}{P(D)}
\end{equation}
where $P(D|M_N)$ is obtained from the fit results, $P(D)$ is a normalization constant, and $P(M_N)$ is the prior, which represents the probability of measuring the observed area given $N$ detected photons in the waveform. This probability is derived from the area distribution of a single photon, which includes the $(22.6 \pm 2.0)\%$ probability of two photoelectrons being produced by a single photon (DPE)~\cite{LopezParedes:2018kzu, Faham-2PEEffect}. The area distributions of more than one photon are generated by the convolution of the area distributions of a single photon. 
After calculating the likelihood score for each $N$, the model with the highest score is selected. Figure~\ref{Fig: N-photon Model Example} presents an example of a 2-photon fit to a waveform.
To limit the bias introduced by DPE, an additional requirement is added in the $N$-photon model: if the reconstructed arrival times of the two photons differ by less than 1 ns, the 2-photon model is rejected and the waveform is treated as a 1-photon case. 

\section{Pulse Shape Discrimination} \label{Sec: PSD}
To explore PSD in LZ, the ER and NR pulse shapes were studied using CH$_3$T and DD calibration data, the same calibrations used to define the ER and NR bands in charge-to-light discrimination. In Sec.~\ref{Sec: Calibration}, the ER/NR photon timing spectra are presented. Section~\ref{Sec: Prompt Fraction Optimization} describes the pulse shape discriminator developed to quantify ER/NR pulse shape differences. Section~\ref{Sec: z-correction on PF} introduces the position correction used to extend the applicability of PSD across the entire fiducial volume (FV).

\subsection{ER/NR Photon Timing Spectra} \label{Sec: Calibration}

CH$_3$T and DD sources are used to optimize the pulse shape discriminator and estimate the discrimination power of PSD. Single-scatter calibration events which pass the DQ cuts are selected~\cite{LZ-WS2024Results}. The CH$_3$T events are approximately uniformly distributed along $z$, where $z$ is the height above the cathode grid. The DD events come from 2.45~MeV neutrons produced by a DD generator that enter the TPC at
$z=136$~cm, approximately 10~cm below the liquid surface~\cite{LZ:calibration}. Since the DD events are occurring at the top of the detector, only DD and CH$_3$T events with $z= 100-140$~cm are selected for the PSD analysis. The pulse area ROI for S1 pulses is $3-80$~phd, the same region that is used in the WIMP search analysis published in Ref.~\cite{LZ-SR1Results,LZ-WS2024Results}. Due to the difficulty of accurately characterizing S1 pulse shapes with low photon statistics, only S1 pulses with an area greater than 20~phd are used for optimizing the pulse shape discriminator. In addition, 
the calibration data above 80 phd have limited statistics and are therefore excluded from the optimization.

\begin{figure}[h]
    \centering
    \includegraphics[width=0.5\textwidth]{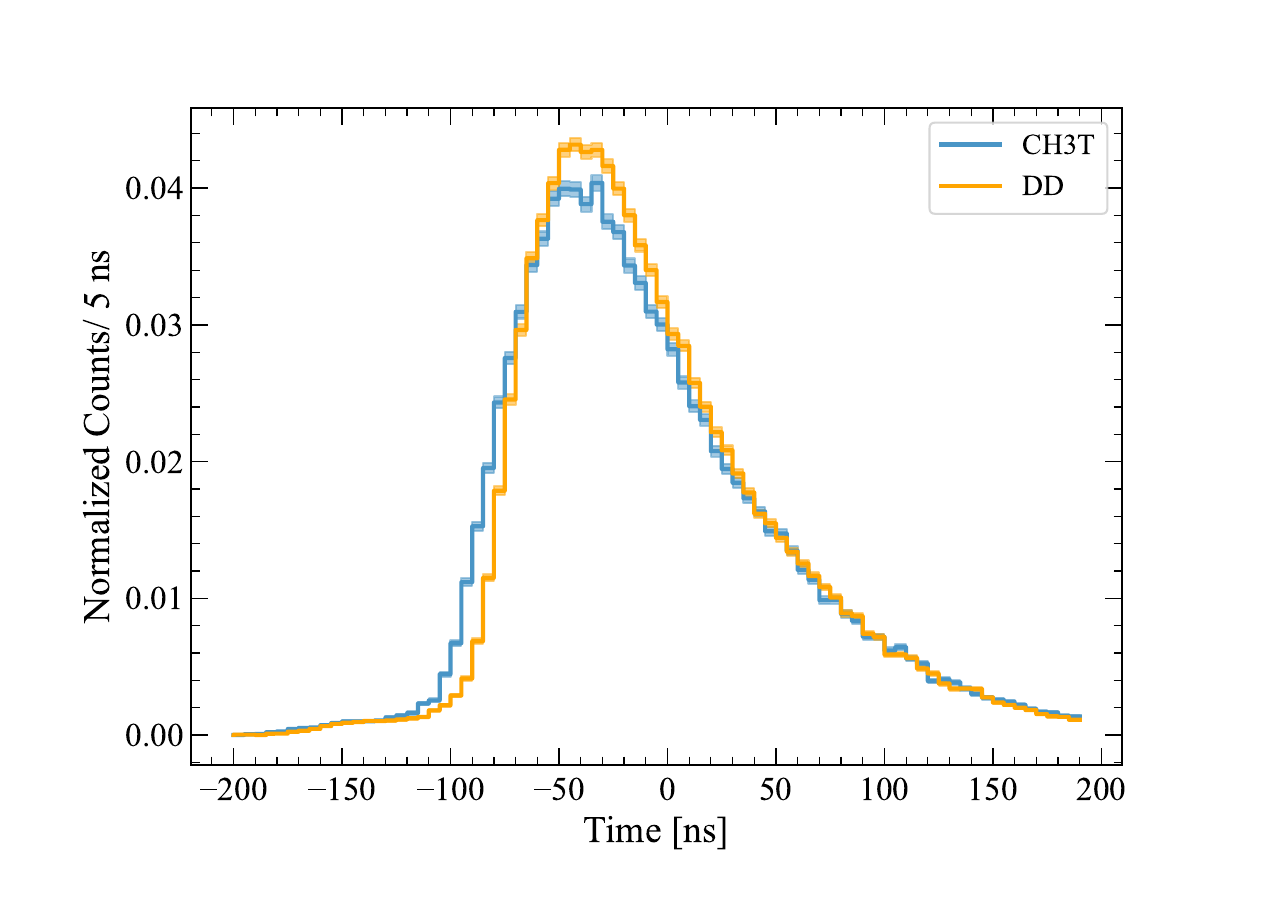}
    \caption{Normalized photon timing distributions for NR (orange) and ER (blue) events with pulse areas between $70-80$~phd and $z= 100-140$~cm. The mean photon arrival time is used to align individual events, defined as $t=0$. Shaded regions represent the statistical uncertainties in each bin. The observed difference between the ER and NR photon timing distributions is exploited to develop the pulse shape discriminator.}
    \label{Fig: Photon Timing Spectrum SR3}
\end{figure}

The $N$-photon model is applied to individual PMT waveforms and used to construct the photon timing distributions for CH$_3$T and DD S1 pulses. The photon timing distributions across different events are aligned based on the mean of the photon arrival times. Since NR and ER events have different ratios of singlet to triplet emission, it is expected that more photons arrive earlier for NR events compared to ER events. An example of the photon timing distributions for events with an area between $70-80$~phd is shown in Fig.~\ref{Fig: Photon Timing Spectrum SR3}. The NR spectrum shows a steeper slope at the rising edge, while the ER spectrum has a longer tail. This difference can be used to discriminate ER and NR events.

\subsection{Tail Fraction Discriminator} \label{Sec: Prompt Fraction Optimization}
To quantify the difference in pulse shape, various quantities can be used, such as full width at half maximum (FWHM), integral/height, prompt/total, etc. In this study, 
the scintillation pulse shape is characterized using the tail fraction (TF), chosen because the optimized parameters obtained from the minimization procedure 
emphasize the importance of tail of the S1 distribution. 
TF is defined as:

\begin{equation}
\label{eq: PF}
\text{TF} = \frac{\sum_{t_0}^{t_1} \text{S1}(t) }{\sum_{t_2}^{t_3} \text{S1}(t) }.
\end{equation}
The summations in the numerator and denominator represent the number of photons within the time windows $t_0-t_1$ and $t_2-t_3$, respectively. Each photon has the same weight, independent of its pulse area. By counting individual photons at low occupancy, the smearing effects of the PMT charge response are eliminated, thereby improving the detector’s resolution. $t_2$ and $t_3$ are fixed at $-200$~ns and 200~ns, respectively, which covers the entire photon timing spectrum. We require that $t_0<t_1$ and optimize both $t_0$ and $t_1$ to minimize the ER leakage in the $50\%$ NR acceptance region, defined above the median of the TF of NR events. The ER leakage is the ratio of the number of ER events that fall into the $50\%$ NR acceptance region and the total number of ER events. 

\begin{figure*}[t!]
    \centering
    \includegraphics[width=0.6\textwidth]{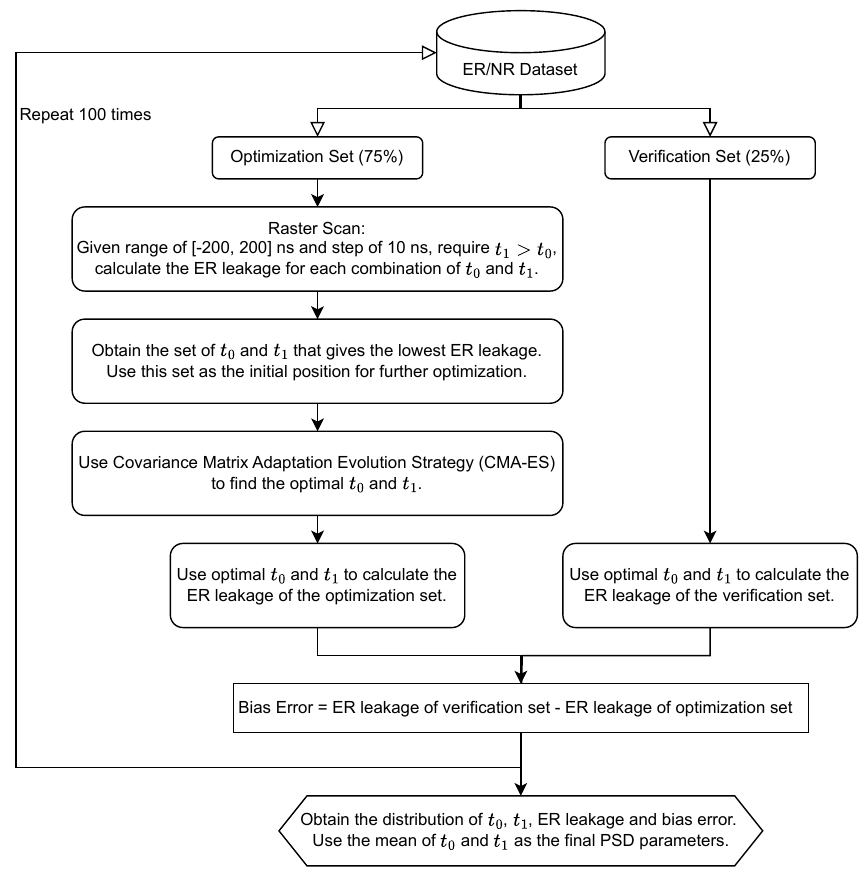}
    \caption{Optimization pipeline for the pulse shape discriminator. The optimization begins with a raster scan over the parameter space, followed by refinement using stochastic gradient descent (SGD). The procedure is repeated 100 times with a $75\%/25\%$ split between optimization and verification. Random sampling is performed in each iteration.}
    \label{Fig: Optimization Pipeline}
\end{figure*}

The optimization is performed separately for each area bin between 20 and 80~phd, with a bin size of 10~phd. The optimization pipeline is illustrated in Fig.~\ref{Fig: Optimization Pipeline}. Both ER and NR datasets are randomly split in a $75\%$ to $25\%$ ratio into an optimization set and a verification set to study the bias. The optimization process is carried out only on the optimization set. The first step of the optimization process is a raster scan of the $t_0-t_1$ parameter space. Within the range from $-200$ to 200~ns, $t_0$ and $t_1$ are varied in steps of 10~ns, and the ER leakage is calculated for each possible combination. The combination of $t_0$ and $t_1$ that minimizes the ER leakage is selected as the starting point for the final optimization. The final optimization is performed using the Covariance Matrix Adaptation Evolution Strategy (CMA-ES)~\cite{CMA-ES}, a stochastic gradient descent (SGD)-based algorithm. The optimized parameters are then applied to calculate ER leakage for both the optimization and verification sets. The difference in ER leakage between these sets is defined as the bias error. This process is repeated 100 times, with random splitting of the events for each iteration. The mean values of the 100 optimized $t_0$'s and $t_1$'s are used as the final PSD parameters. 

\begin{figure}[h]
    \centering
    \includegraphics[width=0.5\textwidth]{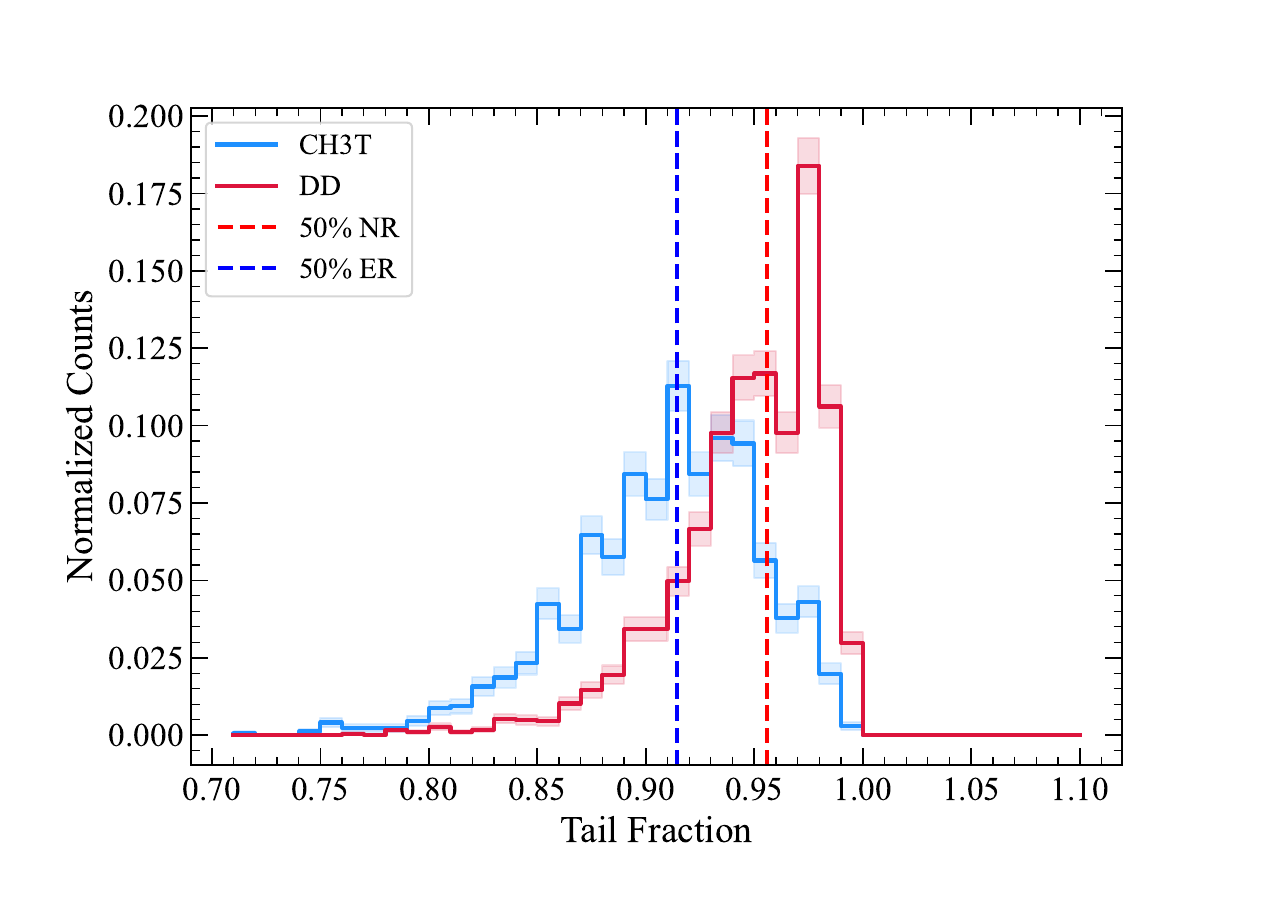}
    \caption{Normalized tail fraction distributions for ER (blue) and NR (red) events with S1$c= 70-80$~phd and $z= 100-140$~cm. The dash lines indicate the median values for ER ($0.914 \pm 0.001$) and NR ($0.956\pm 0.001$). Shaded regions represent the statistical uncertainties in each bin. A clear separation between ER and NR populations is observed in tail fraction space, with an ER leakage of $14.4\pm 1.0 \%$.}
    \label{Fig: PF SR3}
\end{figure}

\begin{figure}[h]
    \centering
    \includegraphics[width=0.5\textwidth]{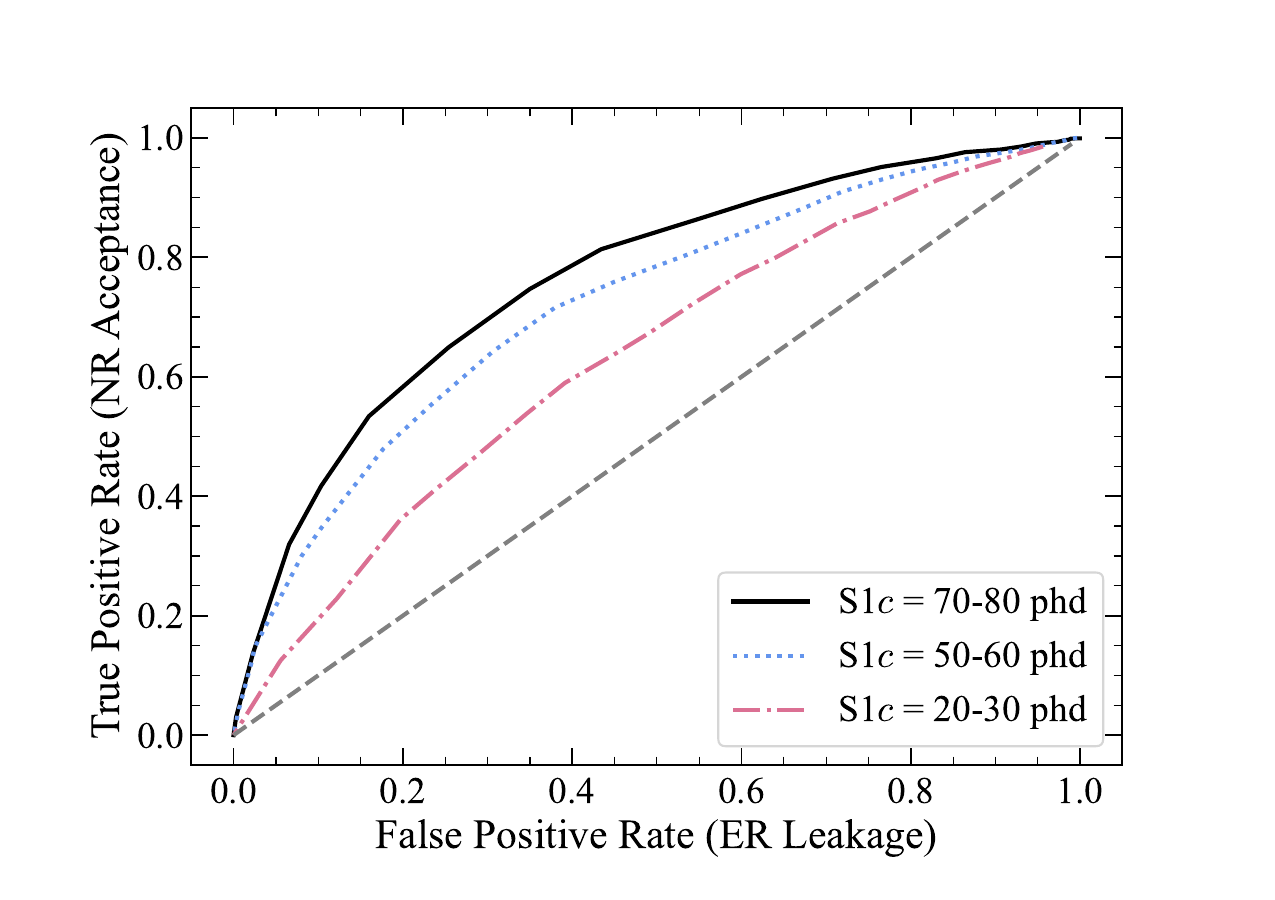}
    \caption{Receiver operating characteristic (ROC) curves for the pulse shape discriminator, evaluated using calibration events with $z= 100-140$~cm, and with S1$c$ of $70-80$~phd (black), $50-60$~phd (blue), and $20-30$~phd (pink). The corresponding areas under the curves are 0.76, 0.72, and 0.59, respectively. The curves are constructed using the optimized tail fraction parameters for the respective bin. The dashed grey line represents the random classifier baseline, for which points above this line indicate performance better than random. }
    \label{Fig: PF ROC}
\end{figure}

\begin{figure}[h]
    \centering
    \includegraphics[width=0.5\textwidth]{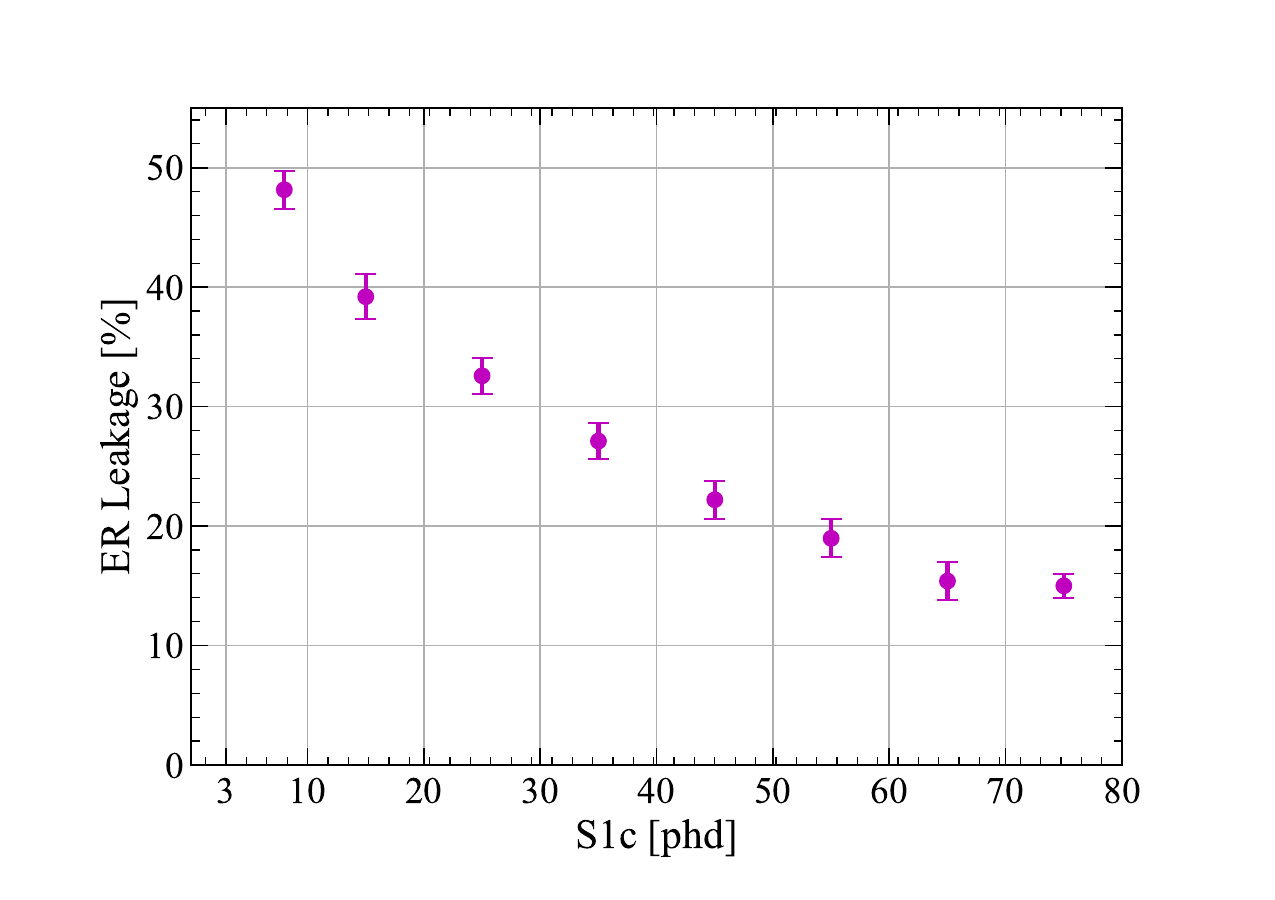}
    \caption{ER leakage in the 50$\%$ NR acceptance region for all area bins within the WS ROI, based on the final optimized TF parameters and calibration events with $z= 100-140$~cm. Error bars represent the bias errors obtained during the optimization process.}
    \label{Fig: ER Leakage}
\end{figure}

The results for the $70-80$~phd area bin are presented as an example, where 
the optimized timing parameters are $t_0 = -71.6 \pm 0.8$~ns and $t_1 = 196 \pm 5$~ns. 
As the photon timing distribution beyond 190~ns contains minimal counts, it is not surprising that $t_1$ has a larger uncertainty. Using the final values of $t_0$ and $t_1$, the TF for all events in this bin is calculated, and the corresponding TF distributions are obtained, shown in Fig.~\ref{Fig: PF SR3}. A clear difference is observed between the ER and NR TF distributions in PSD space. 
The NR distribution appears to the right as a consequence of the choice of $t=0$ used to align the photon timing distributions of different events.
The ER leakage at the 50$\%$ NR acceptance region for this bin is $14\pm 1 \%$. The definition of ER leakage and NR acceptance used in this work is consistent with the concepts of false positive rate (FPR) and true positive rate (TPR), respectively, with 50$\%$ NR acceptance equivalent to a TPR of 0.5. The receiver operating characteristic (ROC) curve, shown in Fig.~\ref{Fig: PF ROC}, characterizes the overall performance of PSD independently of any specific threshold choice. It quantifies the trade-off between FPR and TPR, allowing the selection of an optimal operating point tailored to the specific objective of the analysis. The optimization is performed for all area bins in the range from 20 to 80~phd, with results summarized in Table~\ref{Table: PF Optimization results all}. 
As expected, the ER leakage decreases as the number of photons in the scintillation pulse increases, since more photons more precisely define the photon timing distributions. The bias error shows no significant dependence on photon counts. The final values of $t_0$ and $t_1$ are obtained by calculating their weighted averages. Using these final PSD parameters, the TF for each event is calculated. The ER leakage for each bin is obtained and presented as a function of S1$c$ in Fig.~\ref{Fig: ER Leakage}, where $c$ indicates that a position correction has been applied.
For S1$c>20$~phd, the ER leakage ranges from $15-35\%$ in the 50$\%$ NR acceptance region. In this study, we focused on events below 80 phd, but there is no reason to assume that the optimized $t_0$ and $t_1$ cannot be used for events with S1$c$ above 80 phd.

\begin{table*}[!htbp]
\centering

\caption{Optimization results of the pulse shape discriminator using calibration events with $z= 100-140$~cm, evaluated across S1$c$ bins from 20 to 80~phd. The bias error is obtained based on the averaged difference in ER leakage between the optimization and verification sets over 100 iterations. The ER leakage is computed using all calibration events with $z= 100-140$~cm and passing all DQ cuts, based on the corresponding optimized PSD parameters.  }
\label{Table: PF Optimization results all}
\begin{tabular}{c|c c c c c c }
\hline
S1$c$ Bin [phd] & $t_0$ [ns] & $t_1$ [ns] & ER Leakage [$\%$] & Bias Error [$\%$] & $\#$ of NR events & $\#$ of ER events\\
\hline 

20-30 & -63 $\pm$ 4 & 197 $\pm$ 9 & 32.5 & 1.5 & 5528 & 7044\\

30-40 & -66 $\pm$ 2 & 197 $\pm$ 4 & 26.3 & 1.5 & 3398 & 5753\\

40-50 & -68 $\pm$ 2 & 196 $\pm$ 6 & 23.1 & 1.6 & 2446 & 4615\\

50-60 & -68 $\pm$ 2 & 195 $\pm$ 7 & 19.2 & 1.6 & 2052 & 3531\\

60-70 & -70 $\pm$ 1 & 196 $\pm$ 3 & 15.1 & 1.6 & 2037 & 2630\\
70-80 & -72 $\pm$ 1 & 196 $\pm$ 5 & 14.4 & 1.0 & 2253 & 1720\\
\hline 
Weighted average of $t_0$ and $t_1$ & -69.7 $\pm$ 0.5 & 196 $\pm$ 2 & - & - & 17714 & 25293\\
\hline
\end{tabular}
\end{table*}

\subsection{$z$-correction of the Tail Fraction} \label{Sec: z-correction on PF}
\begin{figure}[h]
    \centering
    \includegraphics[width=0.5\textwidth]{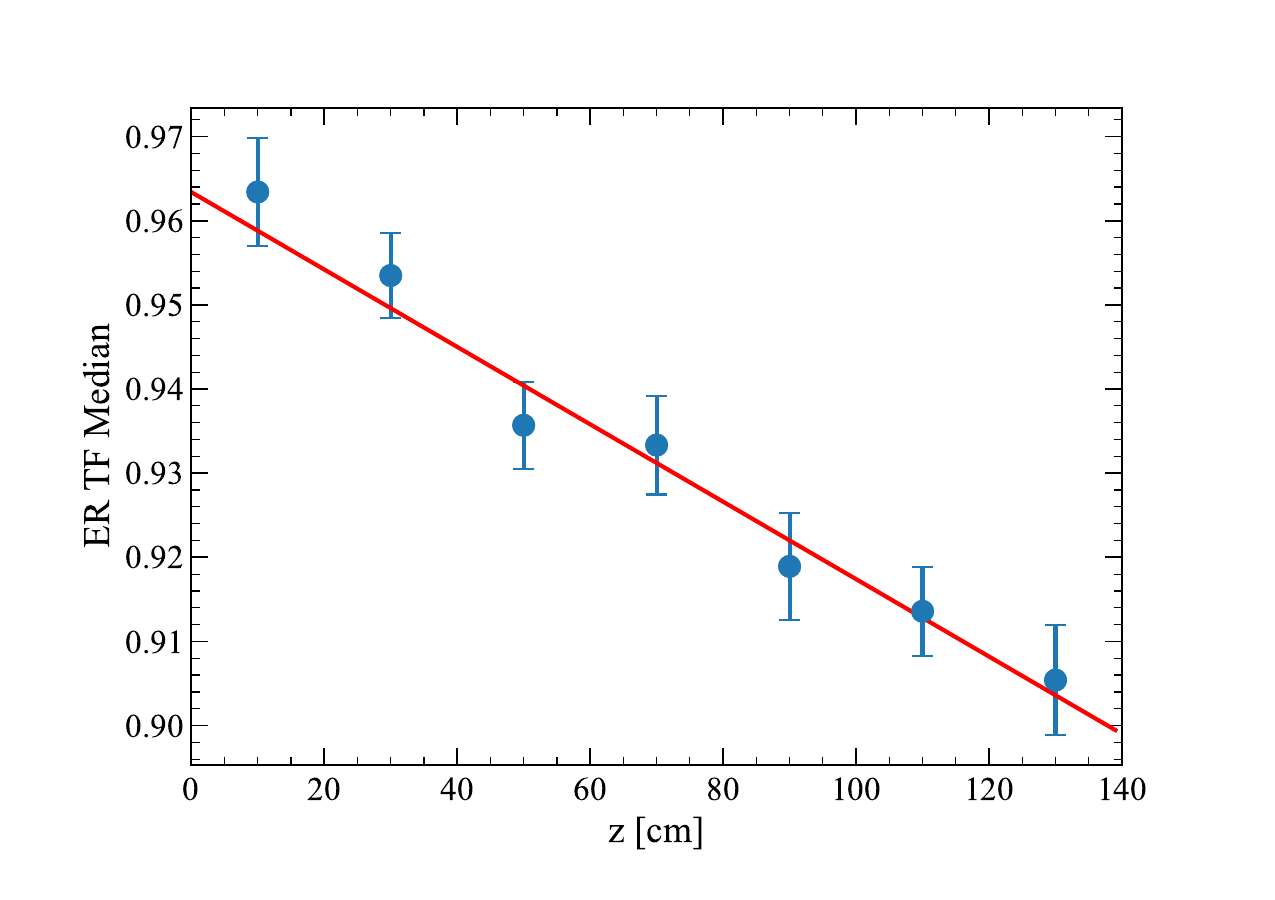}
    \caption{TF medians as function of $z$ for CH$_3$T events with a linear fit (solid red line). The error bars are obtained based on the standard deviation of median. A position-dependent correction, $\Delta \text{TF}(z)$, is derived from this relationship to account for the $z$-dependent variations in TF.  }
    \label{Fig: PF Medians}
\end{figure}

The pulse shape discriminator is developed using calibration events with $z=100-140$~cm due to the space constraint of DD neutron events. WS events are distributed across the entire FV, spanning $z$ $\sim 3-134$~cm. Using CH$_3$T calibration data, we observed that the median of the TF decreases as $z$ increases, as shown in Fig.\ref{Fig: PF Medians}. A linear relationship is observed between the median TF values and $z$. 
To extend PSD across the full FV, a $z$-correction is required. The TF distributions for CH$_3$T events at different radial positions show no observable differences, indicating that only a position correction for the $z$-coordinate is required.

To determine this $z$-dependent correction, we use CH$_3$T events across the entire FV. Since the photon timing spectra show no significant dependence on S1$c$, we select events with a pulse area between 40 and 80~phd. The CH$_3$T events are divided into seven height bins, with $z$ ranging from 0 cm to 140~cm, in 20~cm increments. In each bin, TF values are calculated using the final PSD parameters, described in Sec.~\ref{Sec: Prompt Fraction Optimization}, and the median TF is used as the reference value for correction. 
The correction $\Delta \text{TF}(z)$ is defined as the difference between each bin's median TF and that of the highest bin ($z=120-140$~cm). The corrected tail fraction, $\text{TF}_{\text{cor}}$, is then given by:
\begin{equation}
\label{Eq: PF correction}
    \text{TF}_{\text{cor}} = \text{TF} - \Delta \text{TF}(z).
\end{equation}
A linear fit of $\Delta \text{TF}(z)=az+b$ yields 
$a$~$=$~$ (-4.6$~$\pm 0.5)$~$ \times 10^{-4}$~cm$^{-1}$ and $b = (5.8 \pm 0.4) \times 10^{-2}$. 
This $z$-correction on the tail fraction was not applied in LUX analysis~\cite{Akerib_2018}.

\section{NEST Simulation} \label{Sec: NEST}
The Noble Element Simulation Technique (NEST) is a versatile Monte Carlo-based framework, designed to simulate both the scintillation and ionization responses in noble element detectors~\cite{NEST-general}. NEST is widely used in dark matter and neutrino experiments, and incorporates key parameters such as electric field, temperature, and recombination effects. 

NEST facilitates PSD studies by providing detailed event-level information, including S1$c$, S2$c$, and event position, while excluding PMT and signal processing effects. Simulated ER and NR events from NEST allows the estimation of the performance of PSD under ideal conditions. Comparing NEST and LZ data enables the extraction of key physics parameters of the photon emission model. For example, the recombination time, a critical parameter for noble element detectors, can be determined. This parameter directly affects both the scintillation and ionization yields, which are crucial for defining the detector’s threshold and ER/NR discrimination. Knowing the recombination time improves our understanding of the detector’s performance and informs design choices for future detectors.

\subsection{Photon Timing Model}
Photon timing is defined as the interval between the time of the interaction and the time of detection of the photons by the PMTs. NEST directly simulates the number of photons produced in each interaction and the timing of the individual photons in the S1 pulse. The photon timing model in NEST is composed of the photon emission model and the photon transit model. 

\subsubsection{Photon Emission Model}

The photon emission model is based on the scintillation emission distribution:

\begin{equation}
    P_{\text{emission}}(t) = E_1\frac{e^{-t/\tau_1}}{\tau_1}+(1-E_1)\frac{e^{-t/\tau_3}}{\tau_3}
\end{equation}
where the ratio of singlet emission to triplet emission is given by $E_1/(1-E_1)$. The values of $\tau_1$ and $\tau_3$ are $3.27 \pm 0.66$~ns and $25.89 \pm 0.06$~ns, respectively, and the ratio $E_1/(1-E_1)$ is $0.269 \pm 0.006$ for NR and $0.042 \pm 0.022$ for ER. These values are obtained from the PSD analysis in LUX~\cite{Akerib_2018, NEST-tauRmodel} and should apply for all LXe detectors. 

For photons produced via recombination rather than direct excitation, an additional recombination time must be added. The ratio of direct excitation to recombination follows the built-in NEST model, which differs between ER and NR interactions, as described in Ref.~\cite{Szydagis:2022ikv}. The recombination time used in this work is modeled according to Ref.~\cite{Kubota:1979ugr}:

\begin{equation}
    P_{\text{recombination}}(t) = \frac{\tau_r}{(\frac{t}{\tau_r}+1)^2}
\end{equation}
where $\tau_r$ is the recombination time constant. Various detector conditions, such as drift field, can potentially influence $\tau_r$.  
An adjustment of $\tau_r$ is performed independently for ER and NR to match the photon timing distributions from CH$_3$T and DD calibration events.
The best-fit values for $\tau_r$ are $0.0\pm  0.2$~ns for NR and $3.7 \pm 0.5$~ns for ER. These values are consistent with predictions from the field-dependent recombination model in NEST~\cite{NEST-tauRmodel}. In LUX, $\tau_r$ is negligible due to suppression by the relatively high electric field, which was $\sim180$~V/cm during LUX WS2013 science run~\cite{LUX:2018akb}. 

\subsubsection{Photon Transit Model}
The transport time of photons within the detector depends on several physical properties of the detector internals, including reflection and absorption at the internal surfaces, reflection and transmission at the liquid/gas interface, and absorption and scattering in the liquid. The photon transit model follows the empirical distributions build into NEST~\cite{Akerib_2018}:
\begin{equation}
    P_{\text{transit}}(t) = A\delta(t) + (1-A)\left[\frac{B}{\tau_a}e^{-t/\tau_a}+\frac{1-B}{\tau_b}e^{-t/\tau_b}\right]
    \label{eq:photon transit model}
\end{equation}
where $A$, $B$, $\tau_a$ and $\tau_b$ are parameters obtained by matching NEST to BACCARAT photon transit simulations. BACCARAT is a Geant4-based program to simulate the response of LZ detector~\cite{LZ-simulations}. The first delta function term represents the transit time of photons traveling directly to the PMTs. The second double exponential term represents photons undergoing reflections and scattering within the detector. 
In BACCARAT, photon transit times were simulated by placing a photon source at specific locations in the detector. Two simulations were performed at different radial positions: $r=9.4$~cm and $r=67$~cm. The photon transit time spectra at these two radial positions are compared for each depth bin (10~cm increments), revealing no significant radial dependence. 
The parameters in Eq.\eqref{eq:photon transit model} can be expressed as polynomial functions in $z$, as summarized in Eq.\eqref{Eq:transit parameters}. 
\begin{align}
\label{Eq:transit parameters}
A(z) &= \sum_{i=0}^{2} A_iz^i \quad \quad \quad
B(z) = \sum_{i=0}^{2} B_iz^i \notag\\
\\
\tau_a(z) &= \sum_{i=0}^{3} \alpha_iz^i  \quad \quad \quad
\tau_b(z) = \sum_{i=0}^{3} \beta_iz^i \notag\\ 
\notag
\end{align}
The coefficients of the functions are provided in the Appendix.

A discrepancy was observed between the photon transit time distributions from BACCARAT and the LZ calibration data. 
The difference varies with $z$ and is largest (around 15~ns) at the lowest $z$, decreasing with increasing $z$. 
To remove this $z$-dependency, an additional parameter $k$ was introduced into the photon transit model when matching the NEST model to the LZ calibration data. The parameter $k$ was defined as
\begin{equation}
\label{eq: k(z)}
    k(z)=\frac{z}{2}\times \rho
\end{equation}
where $\rho$ is a uniformly distributed random number between 0 and 1.

\subsection{PSD with NEST}
\begin{figure}[h]
    \centering
    \includegraphics[width=0.5\textwidth]{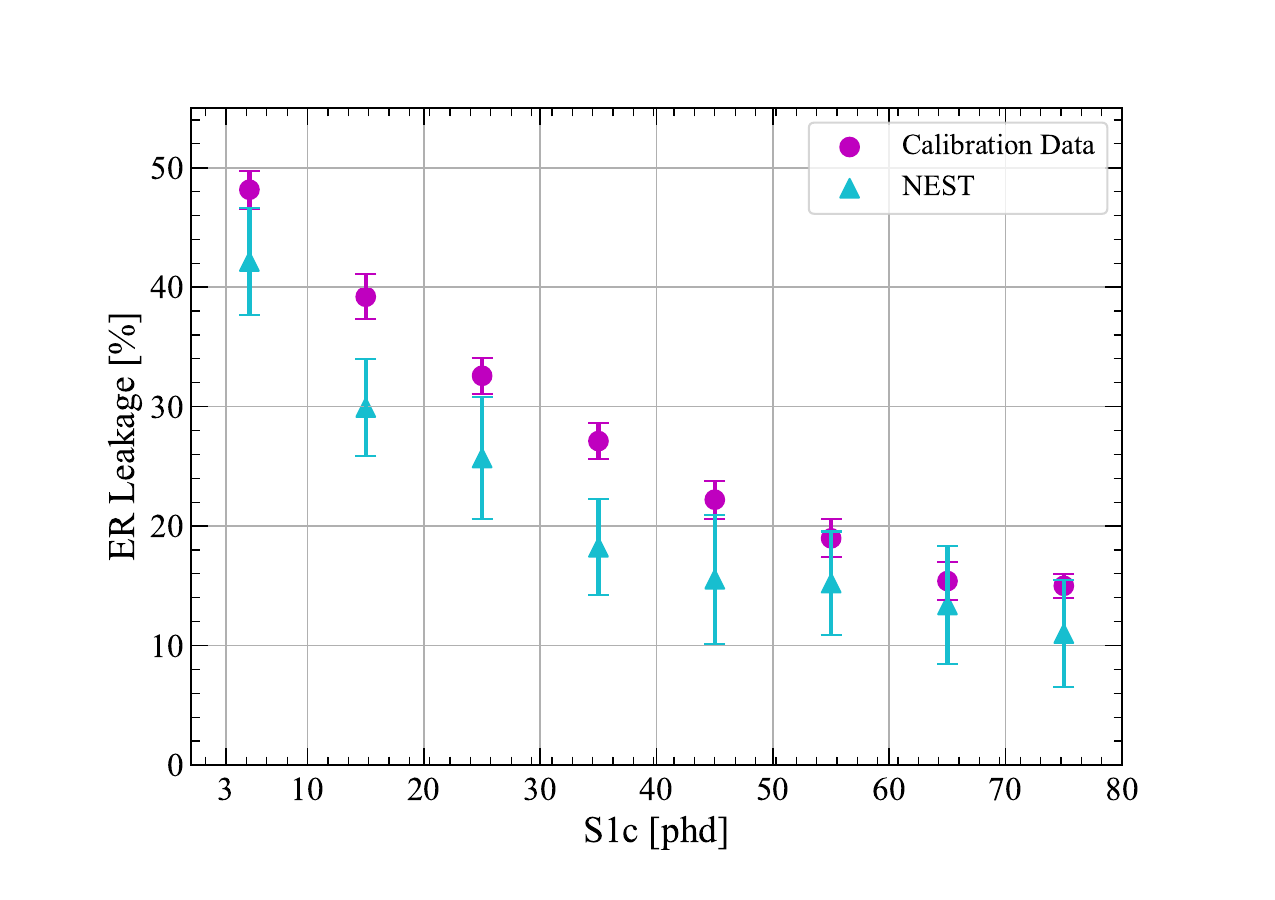}
    \caption{ER leakage in the 50$\%$ NR acceptance region based on NEST-simulated CH$_3$T and DD events (blue), compared with leakage results from calibration data (magenta). Error bars on the NEST results represent the variation in the ER leakage obtained by shifting the optimized parameters in the NEST photon timing model by $\pm 1\sigma$.}
    \label{Fig: NEST ER Leakage}
\end{figure}

NEST-simulated CH$_3$T and DD S1s are used to optimize the pulse shape discriminator, following the same optimization procedure applied to calibration data. The resulting ER leakages in the 50$\%$ NR acceptance region are presented in Fig.~\ref{Fig: NEST ER Leakage}, compared with the results from calibration events. 
The uncertainties on the NEST-based leakage rates are estimated by varying the optimized value of parameters in the photon timing model by $\pm 1\sigma$ and calculating the difference in leakage. For most area bins, the ER leakages from NEST are systematically lower than what we obtained from the calibration data. This is not a surprise since the uncertainties in the signal processing chain are not incorporated in NEST.

\subsection{$z$-correction of the Tail Fraction with NEST}
\begin{figure}[h]
    \centering
    \includegraphics[width=0.5\textwidth]{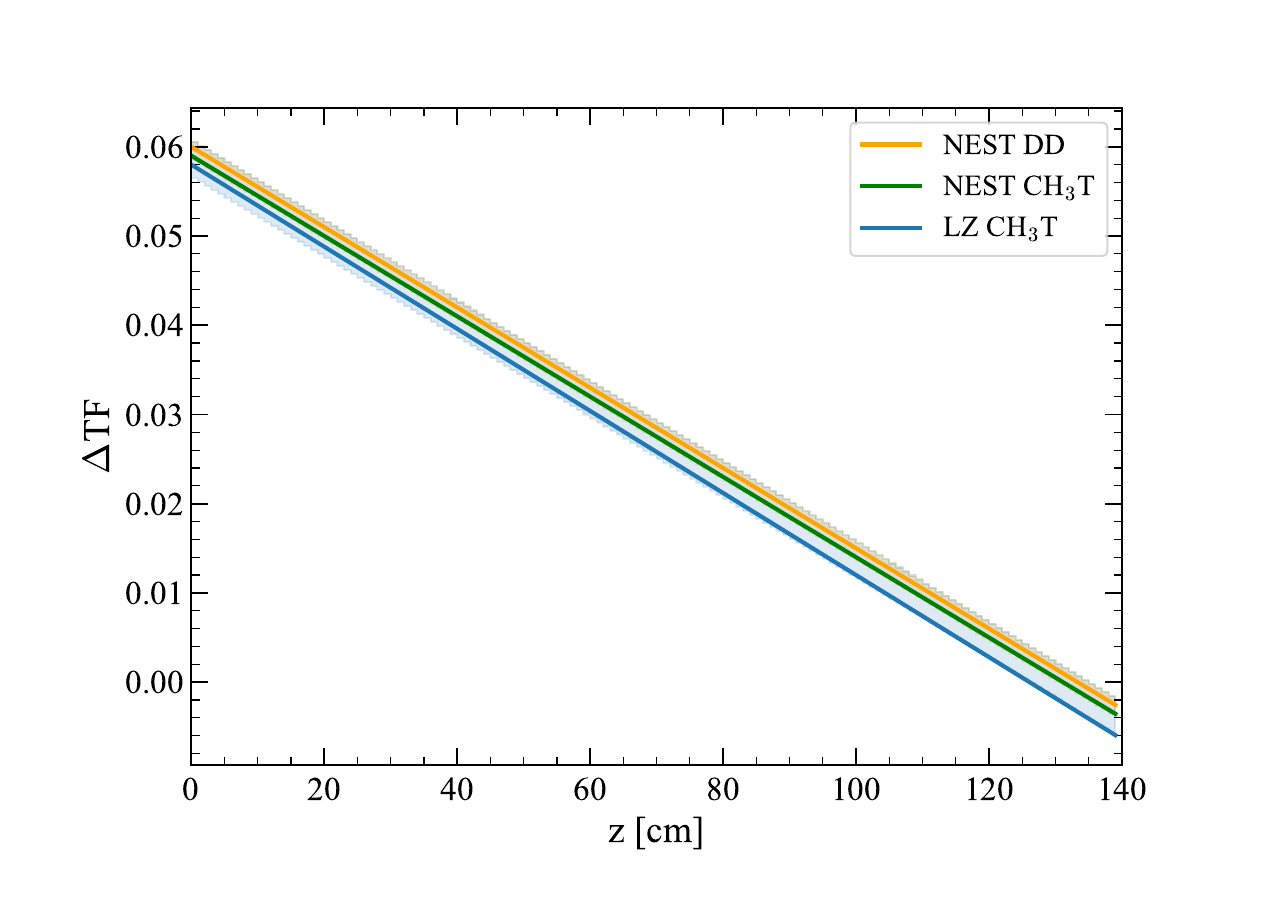}
    \caption{$\Delta \text{TF}(z)$ obtained for NEST events, compared with $\Delta \text{TF}(z)$ obtained for CH$_3$T events. The shaded region represents the 1$\sigma$ uncertainty of the coefficient $b$ in $\Delta \text{TF}(z)$. The position-dependent TF corrections derived from NEST ER, NEST NR, and LZ CH$_3$T events agree within one standard deviation, demonstrating both the consistency between NEST and the LZ datasets and the applicability of the correction to both ER and NR populations.  }
    \label{Fig: NEST PF Correction}
\end{figure}

Since DD events are located in the top third of the detector, direct verification of whether $\Delta \text{TF}(z)$ is the same for NR and ER events is not feasible. To address this, NEST-simulated events are used to study whether $\Delta \text{TF}(z)$ obtained for ER events can be applied for NR events. In NEST, the electric field in the drift region is assumed to be uniform. Although small variations in the drift field can affect the recombination time constant, their impact on ER leakage is estimated to be less than 1$\%$.
NEST-generated CH$_3$T and DD events, with areas between $40-80$~phd, are divided into seven height bins (20~cm per bin), ranging from 0~cm to 140~cm. The TF for each event and the median TF for each bin are calculated. $\Delta \text{TF}(z)$ is determined in the same way as discussed in Sec.~\ref{Sec: z-correction on PF}. As 
shown in Fig.~\ref{Fig: NEST PF Correction}, the linear fit coefficients obtained from NEST DD $\Delta \text{TF}(z)$
agree with those from the NEST CH$_3$T $\Delta \text{TF}(z)$ 
within one standard deviation. Therefore, it is concluded that there is no significant difference in $\Delta \text{TF}(z)$ for ER and NR events. This result is expected, as the photon transit time is unaffected by whether the photon originates from a singlet or triplet molecular state, meaning the $z$-correction is the same for both interaction types. The coefficients of $\Delta \text{TF}(z)$ from NEST and the CH$_3$T calibration data are consistent within one standard deviation. This agreement further supports the assumption that $\Delta \text{TF}(z)$ is applicable regardless of the interaction type or event origin, whether simulated or calibration events. Consequently, the $\Delta \text{TF}(z)$ obtained for the CH$_3$T calibration events is applied to all events.

\subsection{PSD in Charge-suppressed Electron Capture Decays}

Double electron capture (DEC) decays of $^{124}$Xe exhibit a suppressed charge yield compared to standard electron capture (EC) decays~\cite{LZ-Xe124}, causing them to leak into the NR band defined by charge-to-light discrimination. This leakage presents a challenging background for WIMP searches. In the latest LZ search results, $^{124}$Xe DEC was included in the background model and was accounted in the profile likelihood ratio analysis~\cite{LZ-WS2024Results}. PSD is therefore of particular interest for this background, especially since the expected S1$c$ range of these decays (50-80~phd) coincides with the region where PSD provides relatively strong discrimination power.

\begin{figure}[h]
    \centering
    \includegraphics[width=0.5\textwidth]{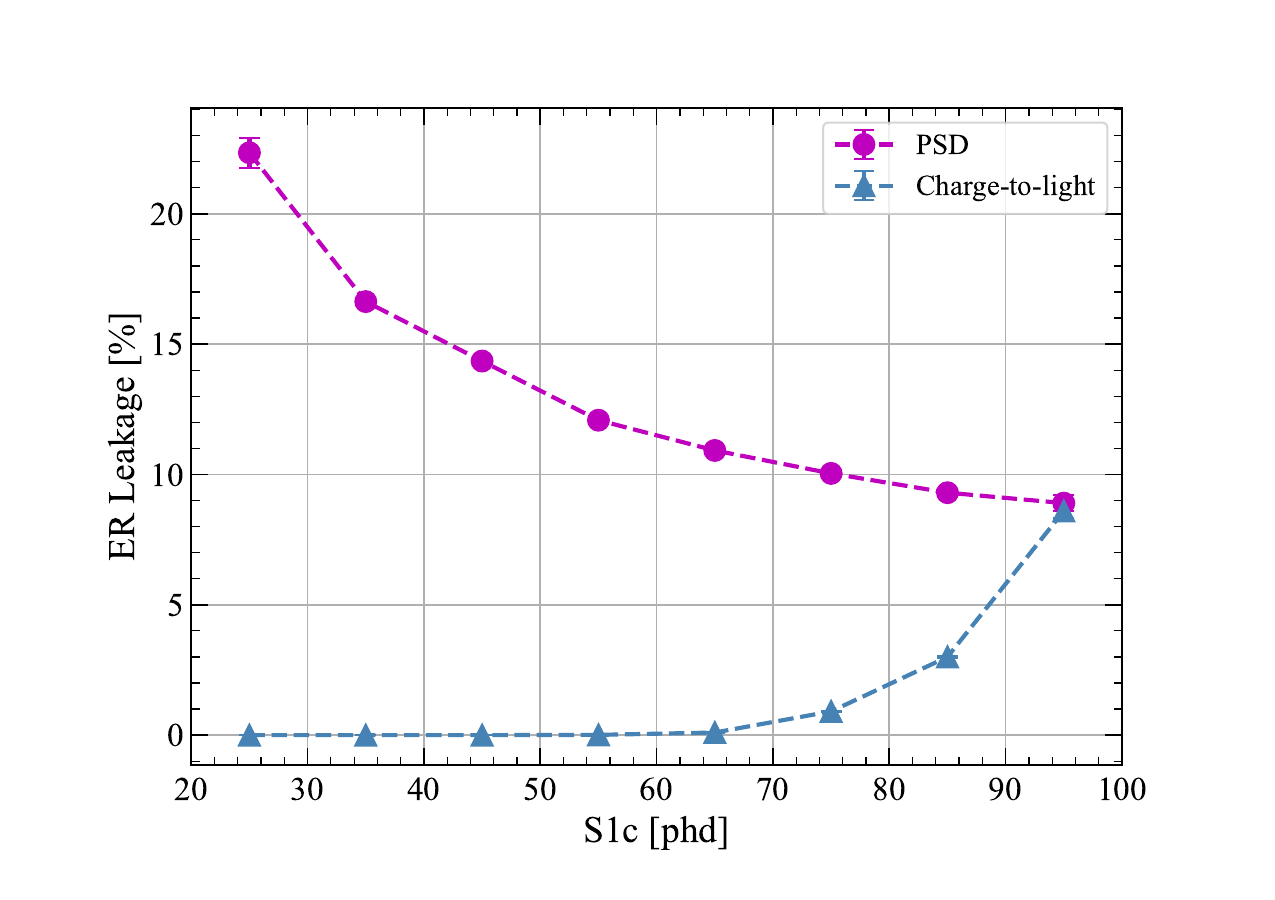}
    \caption{Leakage of $^{124}$Xe DEC events into the 50$\%$ NR acceptance region, based on 10 million NEST-simulated $^{124}$Xe DEC events, using charge-to-light discrimination (blue) and pulse shape discrimination (magenta). Error bars represent Poisson counting uncertainties. The leakage from charge-to-light discrimination increases with S1$c$, while the leakage from PSD decreases as S1$c$ increases.}
    \label{Fig: NEST DEC Leakage}
\end{figure}

Due to the limited statistics available for $^{124}$Xe DEC decays, the NEST photon timing model is used to evaluate the effectiveness of PSD for this background. 
In this study, only LL DEC decays of $^{124}$Xe are considered, serving as an illustrative example of the PSD performance.
To model the S1 photon timing of $^{124}$Xe DEC events, 10 million events were generated using NEST, incorporating the photon timing response described earlier. The leakage of DEC events into the 50$\%$ NR acceptance region is evaluated using both charge-to-light and PSD methods, and is shown in Fig.~\ref{Fig: NEST DEC Leakage}. For S1$c$ values below 70~phd, no leaked events are observed using charge-to-light discrimination alone, indicating that the potential benefit of PSD is limited. Above 70 phd, the ER leakage from the charge-to-light discrimination increases as S1$c$ increases. 
Above 95 phd, PSD is expected to provide better discrimination than charge-to-light, 
making it a valuable tool for background suppression in this energy range.

\section{Two-factor Discrimination} \label{Sec: TFD}

To enhance the discrimination between ER and NR events, a two-factor discrimination (TFD) method was used, combining charge-to-light discrimination and PSD into a unified metric through a linear regression approach. TFD is defined as:
\begin{equation}
    \label{eq: two-factor}
    \text{TFD} = -1 \times \text{CTL}+w(\text{S1}c)\times \text{TF}_{\text{cor}}
\end{equation}
where CTL represents the charge-to-light discrimination output described in detail in the next section, $\text{TF}_{\text{cor}}$ represents the PSD output, and $w(\text{S1}c)$ is the weight assigned to PSD. In this model, the weight of CTL is fixed at one to simplify the regression, reducing the number of free parameters from two to one. This simplification is justified, as the absolute value of the CTL weight only affects the position of the ER/NR distributions in the TFD space, with the key factor being the ratio between the weights of CTL and $\text{TF}_{\text{cor}}$. The negative sign preceding CTL is introduced because ER events have a larger charge yield compared to NR events, while NR events exhibit higher TF values. The negative sign ensures that the NR distribution consistently appears to the right of the ER distribution in the TFD space. Since PSD performance improves with larger S1 pulses, $w$ is expected to depend on S1$c$. During optimization, $w(\text{S1}c)$ is tuned to maximize the ER background rejection power of the TFD.

\subsection{Charge-to-light Discrimination}
\begin{figure}[h]
    \centering
    \includegraphics[width=0.5\textwidth]{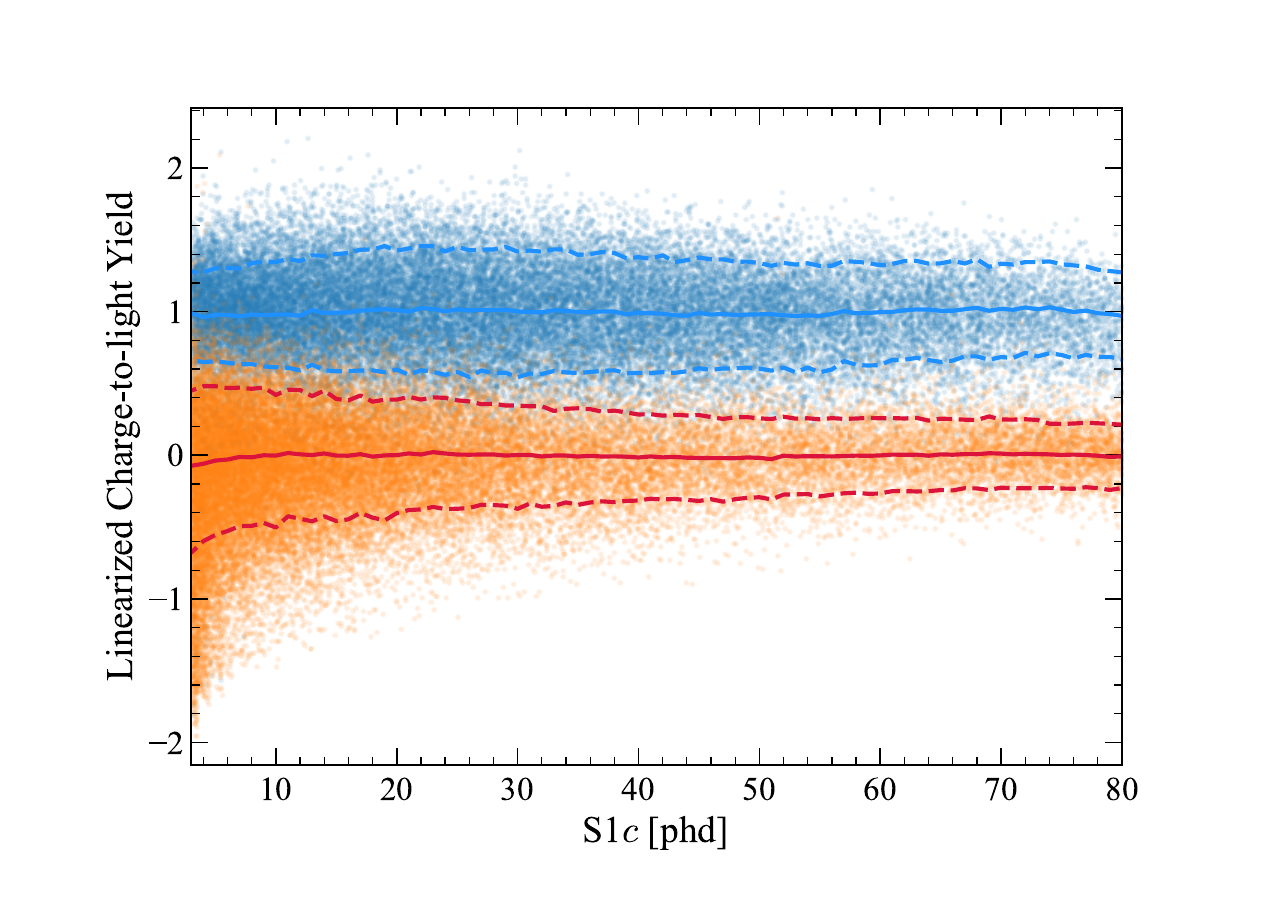}
    \caption{CH$_3$T (blue) and DD (orange) events after linearization. The solid blue and red lines indicate the medians of the ER and NR distributions, respectively, after linearization. The dashed lines represent the 10$\%$ and 90$\%$ quantiles following the linearization process.}
    \label{Fig: CTL}
\end{figure}

The charge-to-light discrimination 
relies on the separation of the energy-dependent ER and NR event distributions in $log_{10}(\text{S2}c)$ versus S1$c$ space~\cite{LZ-SR1Results}. 
A linearization is applied to convert charge-to-light discrimination from 2D to 1D. 
The 1D charge-to-light discriminator (CTL) is defined as:
\begin{equation}
    \text{CTL} = a(\text{S1}c)\times log_{10}(\text{S2}c)+b(\text{S1}c)
\end{equation}
where $a$ and $b$ are polynomial functions of S1$c$,
defined
such that DD and CH$_3$T events are centered around 0 and 1, respectively. 
Figure~\ref{Fig: CTL} shows the calibrations events after linearization. We verified that the ER leakage of the charge-to-light discrimination remains the same before and after linearization.

\subsection{TFD Optimization}
To optimize the two-factor discriminator, CH$_3$T and DD calibration events were used. Events with S1$c$ between 20~and 80~phd are selected. The number of CH$3$T events that leak into the 50$\%$ NR acceptance region based on charge-to-light discrimination alone ranges from $0-10$ across different area bins, as shown in Table~\ref{tab:TFD comparison}. In LUX (Ref.~\cite{Akerib_2018,Akerib_2020}), the improvement of TFD was assessed by quantifying the reduction in the number of leaked events within the 50$\%$ NR acceptance region compared to using charge-to-light discrimination alone. However, the small number of leaked events in LZ leads to a biased optimization and evaluation of TFD performance when relying solely on this statistic. 
An alternative method was introduced to evaluate the performance of TFD by quantifying the separation between the ER and NR distributions. This separation is expressed as the difference between the medians of the ER and NR distributions in the TFD space, $M_{\text{ER}}$ and $M_{\text{NR}}$, over the standard deviation of the ER distribution, $\sigma_{\text{ER}}$:
\begin{equation}
    \label{eq: TFD z-value}
    Z_{\text{score}} = \frac{M_{\text{NR}}-M_{\text{ER}}}{\sigma_{\text{ER}}}.
\end{equation}
This approach, previously used in experiments such as Xenon10~\cite{Dahl:2009nta} and various other PSD studies~\cite{Leland:2024ezs}, is particularly effective when calibration statistics are limited. The $Z_{\text{score}}$ is valid regardless of the shapes of the ER/NR distributions. When the ER distribution is Gaussian-like, the corresponding p-value can be derived from the $Z_{\text{score}}$~\cite{Turner2013}. Here the p-value is interpreted as the Gaussian tail probability, reflecting the expected Gaussian tail leakage of ER events, effectively the false positive rate, and providing an analytic measure of the discrimination power. 
The weight $w$ in Eq.\eqref{eq: two-factor} is optimized by maximizing the $Z_{\text{score}}$. TFD was optimized using CH$_3$T and DD calibration events. These events were divided into 12 bins with a bin size of 5~phd to determine the S1$c$-dependence of $w$. The optimization resulted in a linear relationship: $w(\text{S1}c) = a\times \text{S1}c + b$, where $a=0.0020 \pm 0.0003$~phd$^{-1}$ and $b=0.85\pm0.02$. This function was then used to compute the TFD values for all CH$_3$T events within the FV and for all DD events.

\subsection{Results}
To assess the effectiveness of TFD, its performance was compared to that of charge-to-light discrimination alone by using the $Z_{\text{score}}$ and the corresponding p-values. These results are summarized in Table~\ref{tab:TFD comparison}.
\begin{figure*}[ht!]
    \centering
    \includegraphics[width=0.9\textwidth]{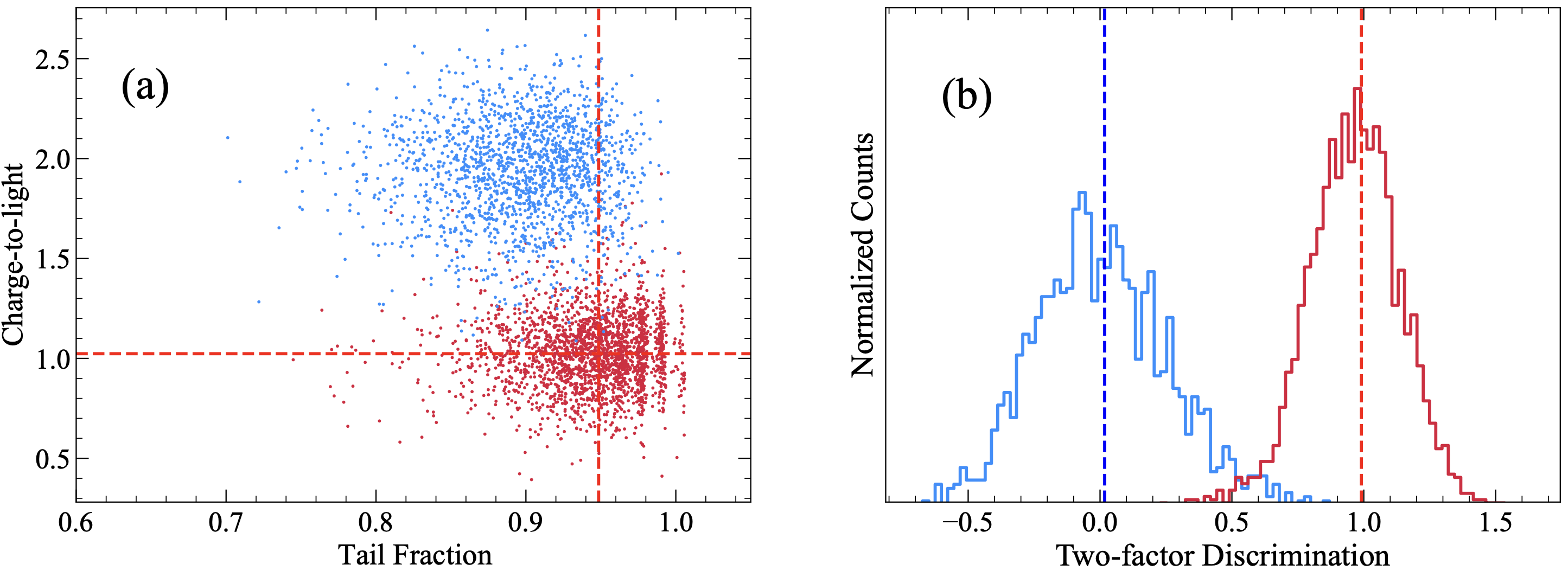}
    \caption{Example of two-factor discrimination (TFD) results using CH$_3$T (blue) and DD (red) calibration events in the S1$c$ range of $70-80$~phd. (a) Events plotted in charge-to-light versus tail fraction space; dashed red lines indicate the NR medians in charge-to-light or PSD dimensions. 
    (b) Normalized TFD distributions for ER and NR events, with dash lines representing the respective medians. In this bin, using TFD enhances the separation between ER and NR distributions, reducing the corresponding p-value by a factor of two, from $1.3 \times 10^{-4}$ to $0.6 \times 10^{-4}$, compared to charge-to-light discrimination. }
    \label{Fig: Example TFD}
\end{figure*}
Bootstrap resampling with replacement was used to estimate the uncertainties of the $Z_{\text{score}}$. 
The improvement in $Z_{\text{score}}$ from CTL to TFD is statistically significant, under the assumption that only statistical uncertainties are considered.
Across all bins, the separation between the ER and NR distributions improves with increasing S1$c$. This improvement is particularly significant in terms of the p-value. 
Figure~\ref{Fig: Example TFD} shows an example using CH$_3$T and DD events with an S1$c$ between 70 and 80~phd. In Fig.~\ref{Fig: Example TFD}(a), the calibration events are plotted in charge-to-light versus PSD space.
In the TFD space, the separation between the ER and NR distributions is $3.83\sigma$, compared to $3.65\sigma$ using charge-to-light alone. The p-value is reduced from $1.3 \times 10^{-4}$ to $0.6 \times 10^{-4}$. Although the p-value with charge-to-light is already very small, TFD further reduces the false positive rate by approximately a factor of 2. For completeness, Table~\ref{tab:TFD comparison} also includes the number of leaked events within the 50$\%$ NR acceptance region for both TFD and charge-to-light discrimination. These results highlight the enhanced ability of TFD to separate ER and NR events and its potential to improve the acceptance, especially for larger S1$c$ signals.

\begin{table*}[!htbp]
    \centering
    \addtolength{\tabcolsep}{5pt}  
    \begin{tabular}{*7c}
\toprule
\hline
S1$c$ [phd] &  \multicolumn{3}{c}{Charge-to-light Discrimination} & \multicolumn{3}{c}{Two-factor Discrimination}\\
\midrule
{}   & $Z_{\text{score}}$   & p-value & leaked events & $Z_{\text{score}}$ & p-value & leaked events\\
20-30   &  3.148 $\pm$ 0.003  & 0.00082 & 10 & 3.174 $\pm$ 0.003 & 0.00075 & 6  \\
30-40   & 3.241 $\pm$ 0.003  & 0.00060 & 7 &  3.325 $\pm$ 0.004 & 0.00044& 6   \\
40-50   & 3.338 $\pm$ 0.004  & 0.00042 & 8 &  3.396 $\pm$ 0.004 & 0.00034& 4   \\
50-60   & 3.442 $\pm$ 0.004  & 0.00029 & 2 &  3.541 $\pm$ 0.005 & 0.00020 & 1   \\
60-70   & 3.507 $\pm$ 0.006  & 0.00023 & 3 &  3.620 $\pm$ 0.006 & 0.00015 & 2   \\
70-80   & 3.651  $\pm$ 0.006 & 0.00013 & 0 &  3.830 $\pm$ 0.007 & 0.00006& 0   \\
\hline
\bottomrule
\end{tabular}
\caption{Summary of TFD performance across S1$c$ bins from 20 to 80 phd, compared to results using charge-to-light discrimination alone, based on CH$_3$T and DD calibration data. The uncertainties on the $Z_{\text{score}}$ are estimated via bootstrap resampling. Given the approximately Gaussian shapes of the ER and NR distributions in the TFD space, the corresponding p-values are derived from the $Z_{\text{score}}$. The ``leaked events'' column represents the number of CH$_3$T events falling within the 50$\%$ NR acceptance region.}
    \label{tab:TFD comparison}
\end{table*}

\section{WIMP search events} \label{Sec: WS}
For the data collected during WS2024, 1,221 events remain after applying the WS DQ selection. Among these, 329 events pass radon tagging~\cite{LZ-WS2024Results}. In this section, PSD and TFD is applied to these 329 events.

\subsection{PSD on WIMP Search Events}
The final PSD parameters and $\Delta \text{TF}$ are used to calculated $\text{TF}_{\text{cor}}$ for each event. 17.4$\%$ of these events fall into the 50$\%$ NR acceptance region. When considering WS events with an S1$c$ above 20 phd, the fraction falling into the 50$\%$ NR acceptance region decreases to 11.4$\%$.

\begin{figure}[h]
    \centering
    \includegraphics[width=0.5\textwidth]{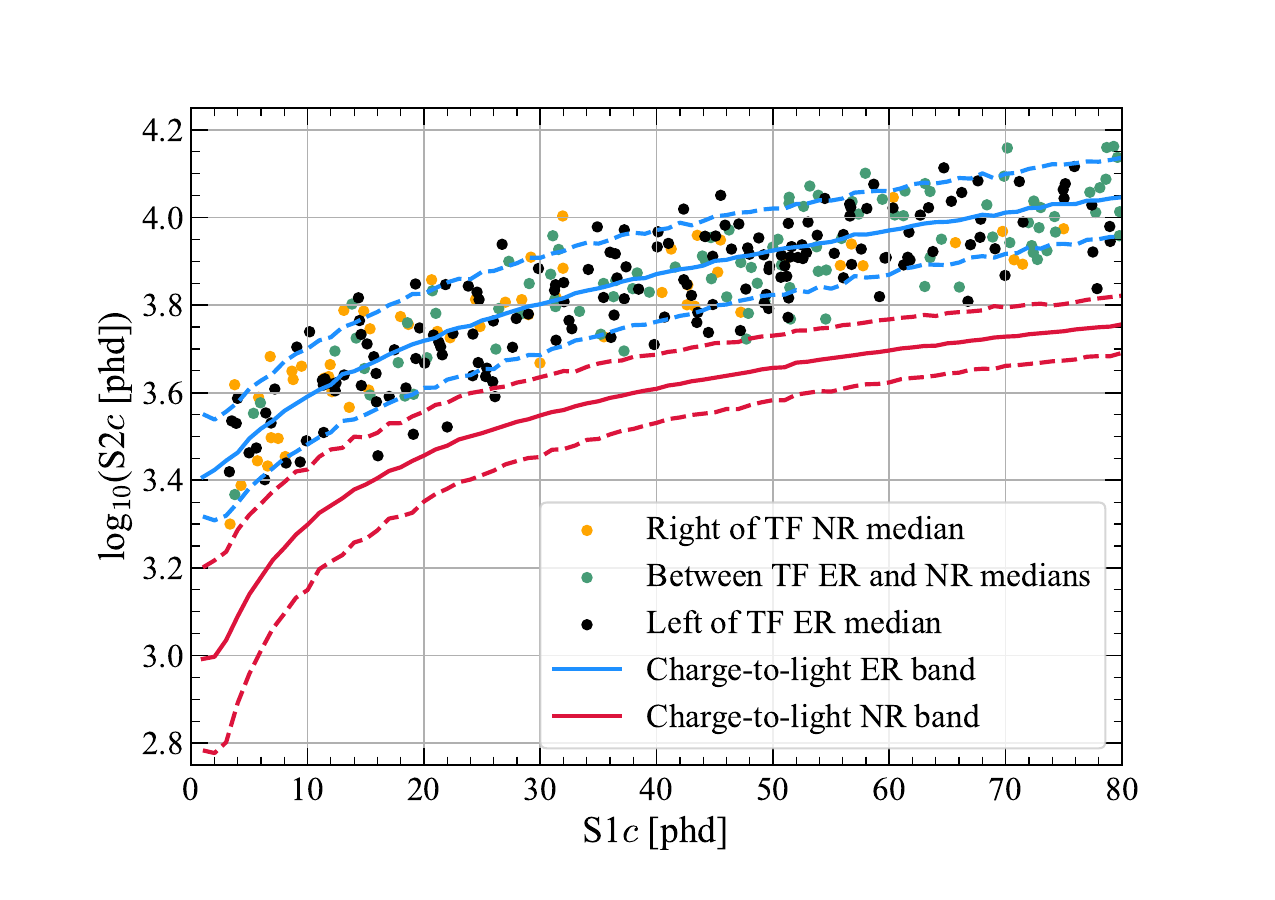}
    \caption{WS2024 events shown in the $log_{10}$S2$c$ versus S1$c$ space, color-coded by pulse shape discriminator (PSD). All four events located within the charge-to-light NR band are classified as ER events based on their tail fraction values.}
    \label{Fig: WS in S1-logS2 space}
\end{figure}

In Fig.~\ref{Fig: WS in S1-logS2 space}, WS events in the $log_{10}$(S2$c$) versus S1$c$ space are color-coded: orange for events within the 50$\%$ NR acceptance region, green for those between the ER and NR medians, and black for those to the left of the ER median. Among events in the charge-to-light NR band, 
all four remaining events 
are outside the PSD NR acceptance region. 

\subsection{TFD on WIMP Search Events}
\begin{figure}
    \centering
    \includegraphics[width=0.5\textwidth]{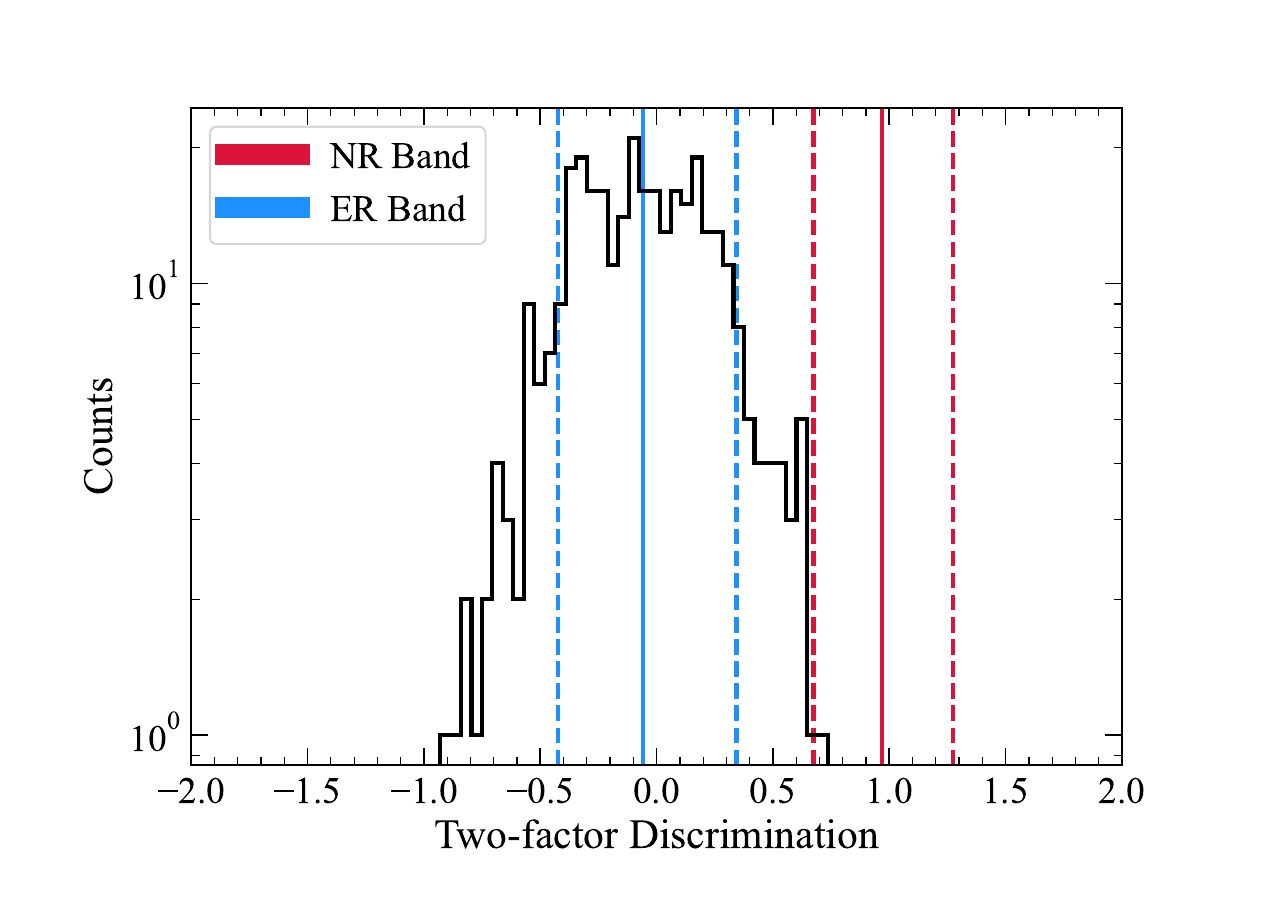}
    \caption{WS2024 events passing radon tagging shown in TFD space. The $10^{\text{th}}$, $50^{\text{th}}$, and $90^{\text{th}}$ percentiles of ER (blue) and NR (red) distributions are indicated. Within the NR band defined by $10^{\text{th}}$ to $90^{\text{th}}$ percentiles, two events are observed using two-factor discrimination, compared to four events when using charge-to-light discrimination alone. In the region between $50^{\text{th}}$ to $90^{\text{th}}$ percentiles of the NR distribution, no events are observed with either method.}
    \label{Fig: WS in TFD space}
\end{figure}

The TFD value of each WS event was calculated using the optimized weighting factor $w(\text{S1}c)$. The distribution of WS events in the TFD space, after applying the radon-tagging selection, is shown in Fig.~\ref{Fig: WS in TFD space}. Following the same approach as the charge-to-light discrimination, the $10^{\text{th}}$, $50^{\text{th}}$, and $90^{\text{th}}$ percentiles of the ER and NR TFD distributions were used to define the corresponding bands in the TFD space. These percentiles-based bands are shown in Fig.~\ref{Fig: WS in TFD space}. 
Two WS events fall within the $10^{\text{th}}$ to $90^{\text{th}}$ percentile NR band in TFD space, compared to four events using charge-to-light discrimination alone. In the $50^{\text{th}}$ to $90^{\text{th}}$ percentile NR region, the number of events remains zero with either method.

\section{Conclusion} \label{Sec: conclusion}
In this paper, the application of PSD in LZ is explored, and an analysis framework is described. To obtain precise photon timing distributions, the relative timing offsets between PMTs were measured, and an $N$-photon model was developed to extract photon timing information from the individual PMT waveforms. PSD was optimized using calibration events. A position correction was applied to the discriminator to extend PSD applicability across the entire fiducial volume. PSD yields an ER leakage between $15$ and $35 \%$ for S1$c>20$~phd. The result shows the power of PSD in a LXe detector of LZ's scale and demonstrates the feasibility of using PSD in future LXe experiments. The photon timing model in NEST was tuned to match the calibration data, validating the photon emission parameters in LXe, and providing an estimate of the best-possible performance of PSD. 

NEST was also used to study the effectiveness of PSD on $^{124}$Xe DEC decays, a background with suppressed charge yields that can leak into the NR band of CTL and can thus be a challenging background to LXe detectors. PSD was shown to reduce this leakage and outperforms CTL above 95 phd, offering a complementary handle on backgrounds that are particularly concerning for the next-generation LXe detector. 

PSD was combined with CTL through a linear regression approach to construct a TFD, which was optimized to maximize the ER/NR separation in TFD space. TFD improved the ER/NR separation relative to CTL separation alone, with the false positive rate being reduced by up to a factor of two. Application of PSD and TFD to WS2024 events provided new insight into the nature of events in the CTL NR band, with results consistent with the null result of LZ. PSD can play an even more powerful role under lower drift-field operating conditions, where increased recombination further enhances timing differences between ER and NR events.

\section*{ACKNOWLEDGMENTS}

The research supporting this work took place in part at the Sanford Underground Research Facility (SURF) in Lead, South Dakota. Funding for this work is supported by the U.S. Department of Energy, Office of Science, Office of High Energy Physics under Contract Numbers DE-AC02-05CH11231, DE-SC0020216, DE-SC0012704, DE-SC0010010, DE-AC02-07CH11359, DE-SC0015910, DE-SC0014223, DE-SC0010813, DE-SC0009999, DE-NA0003180, DE-SC0011702, DE-SC0010072, DE-SC0006605, DE-SC0008475, DE-SC0019193, DE-FG02-10ER46709, UW PRJ82AJ, DE-SC0013542, DE-AC02-76SF00515, DE-SC0018982, DE-SC0019066, DE-SC0015535, DE-SC0019319, DE-SC0025629, DE-SC0024114, DE-AC52-07NA27344, DE-SC0012447. This research was also supported by U.S. National Science Foundation (NSF); the UKRI’s Science $\&$ Technology Facilities Council under award numbers ST/W000490/1, ST/W000482/1, ST/W000636/1, ST/W000466/1, ST/W000628/1, ST/W000555/1, ST/W000547/1, ST/W00058X/1, ST/X508263/1, ST/V506862/1, ST/X508561/1, ST/V507040/1 , ST/W507787/1, ST/R003181/1, ST/R003181/2,  ST/W507957/1, ST/X005984/1, ST/X006050/1; Portuguese Foundation for Science and Technology (FCT) under award numbers PTDC/FIS-PAR/2831/2020; the Institute for Basic Science, Korea (budget number IBS-R016-D1); the Swiss National Science Foundation (SNSF) under award number 10001549. This research was supported by the Australian Government through the Australian Research Council Centre of Excellence for Dark Matter Particle Physics under award number CE200100008. We acknowledge additional support from the UK Science $\&$ Technology Facilities Council (STFC) for PhD studentships and the STFC Boulby Underground Laboratory in the U.K., the GridPP~\cite{GridPP:2006wnd, Britton:2009ser} and IRIS Collaborations, in particular at Imperial College London and additional support by the University College London (UCL) Cosmoparticle Initiative, and the University of Zurich. We acknowledge additional support from the Center for the Fundamental Physics of the Universe, Brown University. K.T. Lesko acknowledges the support of the Royal Society, the Wolfson Foundation, Brasenose College, and the University of Oxford. The LZ Collaboration acknowledges the key contributions of Dr. Sidney Cahn, Yale University, in the production of calibration sources. This research used resources of the National Energy Research Scientific Computing Center, a DOE Office of Science User Facility supported by the Office of Science of the U.S. Department of Energy under Contract No. DE-AC02-05CH11231. We gratefully acknowledge support from GitLab through its GitLab for Education Program. The University of Edinburgh is a charitable body, registered in Scotland, with the registration number SC005336. The assistance of SURF and its personnel in providing physical access and general logistical and technical support is acknowledged. We acknowledge the South Dakota Governor's office, the South Dakota Community Foundation, the South Dakota State University Foundation, and the University of South Dakota Foundation for use of xenon. We also acknowledge the University of Alabama for providing xenon. For the purpose of open access, the authors have applied a Creative Commons Attribution (CC BY) license to any Author Accepted Manuscript version arising from this submission. Finally, we respectfully acknowledge that we are on the traditional land of Indigenous American peoples and honor their rich cultural heritage and enduring contributions. Their deep connection to this land and their resilience and wisdom continue to inspire and enrich our community. We commit to learning from and supporting their effort as original stewards of this land and to preserve their cultures and rights for a more inclusive and sustainable future.

\bibliography{Reference}

@article{Hogenbirk_2018,
   title={Precision measurements of the scintillation pulse shape for low-energy recoils in liquid xenon},
   volume={13},
   ISSN={1748-0221},
   url={http://dx.doi.org/10.1088/1748-0221/13/05/P05016},
   DOI={10.1088/1748-0221/13/05/p05016},
   number={05},
   journal={JINST},
   publisher={IOP Publishing},
   author={Hogenbirk, E. and Aalbers, J. and Breur, P.A. and Decowski, M.P. and Teutem, K. van and Colijn, A.P.},
   year={2018},
   month=may, pages={P05016–P05016} }

@article{Namwongsa_2017,
doi = {10.1088/1748-0221/12/04/P04019},
url = {https://dx.doi.org/10.1088/1748-0221/12/04/P04019},
year = {2017},
month = {apr},
publisher = {},
volume = {12},
number = {04},
pages = {P04019},
author = {P. Namwongsa and A. Banjongkan and X. Chen and K.L. Giboni and X. Ji and C. Kobdaj and H. Kusano and Y. Yupeng},
title = {Electron Recoil rejection by decay time measurement in large liquid Xenon detectors},
journal = {JINST},
}

@article{Kwong_2010,
   title={Scintillation pulse shape discrimination in a two-phase xenon time projection chamber},
   volume={612},
   ISSN={0168-9002},
   url={http://dx.doi.org/10.1016/j.nima.2009.10.106},
   DOI={10.1016/j.nima.2009.10.106},
   number={2},
   journal={Nucl. Instrum. Meth. A},
   publisher={Elsevier BV},
   author={Kwong, J. and Brusov, P. and Shutt, T. and Dahl, C.E. and Bolozdynya, A.I. and Bradley, A.},
   year={2010},
   month=jan, pages={328–333} }

@article{AKIMOV2002245,
    author = "Akimov, D. and others",
    title = "{Measurements of scintillation efficiency and pulse shape for low-energy recoils in liquid xenon}",
    eprint = "hep-ex/0106042",
    archivePrefix = "arXiv",
    reportNumber = "SHEF-HEP-00-4",
    doi = "10.1016/S0370-2693(01)01411-3",
    journal = "Phys. Lett. B",
    volume = "524",
    pages = "245--251",
    year = "2002"
}

@article{Mount:2017qzi,
    author = "Mount, B. J. and others",
    title = "{LUX-ZEPLIN (LZ) Technical Design Report}",
    eprint = "1703.09144",
    archivePrefix = "arXiv",
    primaryClass = "physics.ins-det",
    reportNumber = "LBNL-1007256, FERMILAB-TM-2653-AE-E-PPD",
    month = "3",
    year = "2017"
}

@article{PhysRevB.20.3486,
  title = {Dynamical behavior of free electrons in the recombination process in liquid argon, krypton, and xenon},
  author = {Kubota, Shinzou and Hishida, Masahiko and Suzuki, Masayo and Ruan(Gen), Jian-zhi},
  journal = {Phys. Rev. B},
  volume = {20},
  issue = {8},
  pages = {3486--3496},
  numpages = {0},
  year = {1979},
  month = {Oct},
  publisher = {American Physical Society},
  doi = {10.1103/PhysRevB.20.3486},
  url = {https://link.aps.org/doi/10.1103/PhysRevB.20.3486}
}

@article{Akerib_2020,
    author = "Akerib, D. S. and others",
    collaboration = "LUX",
    title = "{Discrimination of electronic recoils from nuclear recoils in two-phase xenon time projection chambers}",
    eprint = "2004.06304",
    archivePrefix = "arXiv",
    primaryClass = "physics.ins-det",
    doi = "10.1103/PhysRevD.102.112002",
    journal = "Phys. Rev. D",
    volume = "102",
    number = "11",
    pages = "112002",
    year = "2020"
}

@article{PhysRevB.27.5279,
  title = {Effect of ionization density on the time dependence of luminescence from liquid argon and xenon},
  author = {Hitachi, Akira and Takahashi, Tan and Funayama, Nobutaka and Masuda, Kimiaki and Kikuchi, Jun and Doke, Tadayoshi},
  journal = {Phys. Rev. B},
  volume = {27},
  issue = {9},
  pages = {5279--5285},
  numpages = {0},
  year = {1983},
  month = {May},
  publisher = {American Physical Society},
  doi = {10.1103/PhysRevB.27.5279},
  url = {https://link.aps.org/doi/10.1103/PhysRevB.27.5279}
}

@article{Akerib_2018,
    author = "Akerib, D. S. and others",
    collaboration = "LUX",
    title = "{Liquid xenon scintillation measurements and pulse shape discrimination in the LUX dark matter detector}",
    eprint = "1802.06162",
    archivePrefix = "arXiv",
    primaryClass = "physics.ins-det",
    doi = "10.1103/PhysRevD.97.112002",
    journal = "Phys. Rev. D",
    volume = "97",
    number = "11",
    pages = "112002",
    year = "2018"
}

@article{PhysRevC.78.035801,
  title = {Scintillation time dependence and pulse shape discrimination in liquid argon},
  author = {Lippincott, W. H. and Coakley, K. J. and Gastler, D. and Hime, A. and Kearns, E. and McKinsey, D. N. and Nikkel, J. A. and Stonehill, L. C.},
  journal = {Phys. Rev. C},
  volume = {78},
  issue = {3},
  pages = {035801},
  numpages = {12},
  year = {2008},
  month = {Sep},
  publisher = {American Physical Society},
  doi = {10.1103/PhysRevC.78.035801},
  url = {https://link.aps.org/doi/10.1103/PhysRevC.78.035801}
}

@article{FUJII2015293,
title = {High-accuracy measurement of the emission spectrum of liquid xenon in the vacuum ultraviolet region},
journal = {Nucl. Instrum. Meth. A},
volume = {795},
pages = {293-297},
year = {2015},
issn = {0168-9002},
doi = {https://doi.org/10.1016/j.nima.2015.05.065},
url = {https://www.sciencedirect.com/science/article/pii/S016890021500724X},
author = {Keiko Fujii and Yuya Endo and Yui Torigoe and Shogo Nakamura and Tomiyoshi Haruyama and Katsuyu Kasami and Satoshi Mihara and Kiwamu Saito and Shinichi Sasaki and Hiroko Tawara}
}

@article{LZ:calibration,
    author = "Aalbers, J. and others",
    collaboration = "LZ",
    title = "{The design, implementation, and performance of the LZ calibration systems}",
    eprint = "2406.12874",
    archivePrefix = "arXiv",
    primaryClass = "physics.ins-det",
    doi = "10.1088/1748-0221/19/08/P08027",
    journal = "JINST",
    volume = "19",
    number = "08",
    pages = "P08027",
    year = "2024"
}

@article{CMA-ES,
    author = "Hansen, Nikolaus",
    title = "{The CMA Evolution Strategy: A Tutorial}",
    eprint = "1604.00772",
    archivePrefix = "arXiv",
    primaryClass = "cs.LG",
    month = "4",
    year = "2016"
}

@article{NEST-general,
    author = "Szydagis, M. and Barry, N. and Kazkaz, K. and Mock, J. and Stolp, D. and Sweany, M. and Tripathi, M. and Uvarov, S. and Walsh, N. and Woods, M.",
    title = "{NEST: A Comprehensive Model for Scintillation Yield in Liquid Xenon}",
    eprint = "1106.1613",
    archivePrefix = "arXiv",
    primaryClass = "physics.ins-det",
    doi = "10.1088/1748-0221/6/10/P10002",
    journal = "JINST",
    volume = "6",
    pages = "P10002",
    year = "2011"
}

@article{Kubota:1979ugr,
    author = "Kubota, Shinzou and Hishida, Masahiko and Suzuki, Masayo and Ruan(Gen), Jian-zhi",
    title = "{Dynamical behavior of free electrons in the recombination process in liquid argon, krypton, and xenon}",
    doi = "10.1103/PhysRevB.20.3486",
    journal = "Phys. Rev. B",
    volume = "20",
    number = "8",
    pages = "3486",
    year = "1979"
}

@article{NEST-tauRmodel,
    author = "Mock, Jeremy and Barry, Nichole and Kazkaz, Kareem and Szydagis, Matthew and Tripathi, Mani and Uvarov, Sergey and Woods, Michael and Walsh, Nicholas",
    title = "{Modeling Pulse Characteristics in Xenon with NEST}",
    eprint = "1310.1117",
    archivePrefix = "arXiv",
    primaryClass = "physics.ins-det",
    doi = "10.1088/1748-0221/9/04/T04002",
    journal = "JINST",
    volume = "9",
    pages = "T04002",
    year = "2014"
}

@article{LZ-simulations,
    author = "Akerib, D. S. and others",
    collaboration = "LZ",
    title = "{Simulations of Events for the LUX-ZEPLIN (LZ) Dark Matter Experiment}",
    eprint = "2001.09363",
    archivePrefix = "arXiv",
    primaryClass = "physics.ins-det",
    reportNumber = "FERMILAB-PUB-20-401-AE",
    doi = "10.1016/j.astropartphys.2020.102480",
    journal = "Astropart. Phys.",
    volume = "125",
    pages = "102480",
    year = "2021"
}

@article{Faham-2PEEffect,
    author = "Faham, C. H. and Gehman, V. M. and Currie, A. and Dobi, A. and Sorensen, P. and Gaitskell, R. J.",
    title = "{Measurements of wavelength-dependent double photoelectron emission from single photons in VUV-sensitive photomultiplier tubes}",
    eprint = "1506.08748",
    archivePrefix = "arXiv",
    primaryClass = "physics.ins-det",
    doi = "10.1088/1748-0221/10/09/P09010",
    journal = "JINST",
    volume = "10",
    number = "09",
    pages = "P09010",
    year = "2015"
}

@article{LZ-SR1Results,
    author = "Aalbers, J. and others",
    collaboration = "LZ",
    title = "{First Dark Matter Search Results from the LUX-ZEPLIN (LZ) Experiment}",
    eprint = "2207.03764",
    archivePrefix = "arXiv",
    primaryClass = "hep-ex",
    doi = "10.1103/PhysRevLett.131.041002",
    journal = "Phys. Rev. Lett.",
    volume = "131",
    number = "4",
    pages = "041002",
    year = "2023"
}

@article{LZ-WS2024Results,
    author = "Aalbers, J. and others",
    collaboration = "LZ",
    title = "{Dark Matter Search Results from 4.2{\,}{\,}Tonne-Years of Exposure of the LUX-ZEPLIN (LZ) Experiment}",
    eprint = "2410.17036",
    archivePrefix = "arXiv",
    primaryClass = "hep-ex",
    reportNumber = "FERMILAB-PUB-24-0796-V",
    doi = "10.1103/4dyc-z8zf",
    journal = "Phys. Rev. Lett.",
    volume = "135",
    number = "1",
    pages = "011802",
    year = "2025"
}

@article{Rubin:1978kmz,
    author = "Rubin, Vera C. and Ford, Jr., W. Kent and Thonnard, Norbert",
    title = "{Extended rotation curves of high-luminosity spiral galaxies. IV.  Systematic dynamical properties, Sa through Sc}",
    doi = "10.1086/182804",
    journal = "Astrophys. J. Lett.",
    volume = "225",
    pages = "L107--L111",
    year = "1978"
}

@Inbook{Turner2013,
author="Turner, J. Rick",
editor="Gellman, Marc D.
and Turner, J. Rick",
title="Standard Normal (Z) Distribution",
bookTitle="Encyclopedia of Behavioral Medicine",
year="2013",
publisher="Springer New York",
address="New York, NY",
pages="1875--1877",
isbn="978-1-4419-1005-9",
doi="10.1007/978-1-4419-1005-9_1075",
url="https://doi.org/10.1007/978-1-4419-1005-9_1075"
}

@article{Leland:2024ezs,
    author = "Leland, John and Fang, Ming and Pani, Satwik and Venturini, Yuri and Locatelli, Marco and Di Fulvio, Angela",
    title = "{Enabling pulse shape discrimination with commercial ASICs}",
    eprint = "2403.16927",
    archivePrefix = "arXiv",
    primaryClass = "physics.ins-det",
    doi = "10.1016/j.nima.2024.169438",
    journal = "Nucl. Instrum. Meth. A",
    volume = "1064",
    pages = "169438",
    year = "2024"
}

@article{Szydagis:2022ikv,
    author = "Szydagis, M. and others",
    title = "{A review of NEST models for liquid xenon and an exhaustive comparison with other approaches}",
    eprint = "2211.10726",
    archivePrefix = "arXiv",
    primaryClass = "hep-ex",
    doi = "10.3389/fdest.2024.1480975",
    journal = "Front. Detect. Sci. Tech.",
    volume = "2",
    pages = "1480975",
    year = "2024"
}

@article{Holtzman:1989ki,
    author = "Holtzman, Jon A.",
    title = "{Microwave background anisotropies and large scale structure in universes with cold dark matter, baryons, radiation, and massive and massless neutrinos}",
    doi = "10.1086/191362",
    journal = "Astrophys. J. Suppl.",
    volume = "71",
    pages = "1--24",
    year = "1989"
}

@phdthesis{Dahl:2009nta,
    author = "Dahl, Carl Eric",
    title = "{The physics of background discrimination in liquid xenon, and first results from Xenon10 in the hunt for WIMP dark matter}",
    reportNumber = "AAT-3364532",
    school = "Princeton U.",
    year = "2009"
}

@article{LZ-Xe124,
    author = "Aalbers, J. and others",
    collaboration = "LZ",
    title = "{Measurements and models of enhanced recombination following inner-shell vacancies in liquid xenon}",
    eprint = "2503.05679",
    archivePrefix = "arXiv",
    primaryClass = "hep-ex",
    reportNumber = "FERMILAB-PUB-25-0185-V",
    doi = "10.1103/447w-94h3",
    journal = "Phys. Rev. D",
    volume = "112",
    number = "1",
    pages = "012024",
    year = "2025"
}

@article{LZ:2019sgr,
    author = "Akerib, D. S. and others",
    collaboration = "LZ",
    title = "{The LUX-ZEPLIN (LZ) Experiment}",
    eprint = "1910.09124",
    archivePrefix = "arXiv",
    primaryClass = "physics.ins-det",
    reportNumber = "FERMILAB-PUB-19-555-AE-E",
    doi = "10.1016/j.nima.2019.163047",
    journal = "Nucl. Instrum. Meth. A",
    volume = "953",
    pages = "163047",
    year = "2020"
}

@article{LZ:2024bvw,
    author = "Aalbers, J. and others",
    collaboration = "LZ",
    title = "{The data acquisition system of the LZ dark matter detector: FADR}",
    eprint = "2405.14732",
    archivePrefix = "arXiv",
    primaryClass = "physics.ins-det",
    doi = "10.1016/j.nima.2024.169712",
    journal = "Nucl. Instrum. Meth. A",
    volume = "1068",
    pages = "169712",
    year = "2024"
}

@article{LUX:2018akb,
    author = "Akerib, D. S. and others",
    collaboration = "LUX",
    title = "{Results of a Search for Sub-GeV Dark Matter Using 2013 LUX Data}",
    eprint = "1811.11241",
    archivePrefix = "arXiv",
    primaryClass = "astro-ph.CO",
    doi = "10.1103/PhysRevLett.122.131301",
    journal = "Phys. Rev. Lett.",
    volume = "122",
    number = "13",
    pages = "131301",
    year = "2019"
}

@article{LopezParedes:2018kzu,
    author = "L{\'o}pez Paredes, B. and Ara{\'u}jo, H. M. and Froborg, F. and Marangou, N. and Olcina, I. and Sumner, T. J. and Taylor, R. and Tom{\'a}s, A. and Vacheret, A.",
    title = "{Response of photomultiplier tubes to xenon scintillation light}",
    eprint = "1801.01597",
    archivePrefix = "arXiv",
    primaryClass = "physics.ins-det",
    doi = "10.1016/j.astropartphys.2018.04.006",
    journal = "Astropart. Phys.",
    volume = "102",
    pages = "56--66",
    year = "2018"
}

@article{GridPP:2006wnd,
    author = "Faulkner, P. J. W. and others",
    collaboration = "GridPP",
    title = "{GridPP: Development of the UK computing Grid for particle physics}",
    doi = "10.1088/0954-3899/32/1/N01",
    journal = "J. Phys. G",
    volume = "32",
    pages = "N1--N20",
    year = "2006"
}

@article{Britton:2009ser,
    author = "Britton, D. and others",
    title = "{GridPP: the UK grid for particle physics}",
    doi = "10.1098/rsta.2009.0036",
    journal = "Phil. Trans. Roy. Soc. Lond. A",
    volume = "367",
    number = "1897",
    pages = "2447--2457",
    year = "2009"
}

@article{SCENE:2014iyj,
    author = "Cao, H. and others",
    collaboration = "SCENE",
    title = "{Measurement of Scintillation and Ionization Yield and Scintillation Pulse Shape from Nuclear Recoils in Liquid Argon}",
    eprint = "1406.4825",
    archivePrefix = "arXiv",
    primaryClass = "physics.ins-det",
    reportNumber = "FERMILAB-PUB-14-204-AE-E",
    doi = "10.1103/PhysRevD.91.092007",
    journal = "Phys. Rev. D",
    volume = "91",
    pages = "092007",
    year = "2015"
}

@article{Manthos:2023swh,
    author = "Manthos, Ioannis",
    collaboration = "DarkSide-20k",
    title = "{DarkSide-20k: Next generation Direct Dark Matter searches with liquid Argon}",
    eprint = "2312.03597",
    archivePrefix = "arXiv",
    primaryClass = "hep-ex",
    doi = "10.22323/1.449.0113",
    journal = "PoS",
    volume = "EPS-HEP2023",
    pages = "113",
    year = "2024"
}

\clearpage
\newpage
\mbox{~}

\begin{widetext}
    \section*{Appendix}
    \begin{table}[h]
        \centering  
        \begin{tabular}{c c}
        \hline
        \hline
          Parameter   &  Value\\ \hline
    $A_1$   & (1.85 $\pm$ 0.04) $\times 10^{-5}$ cm$^{-2}$\\
    $A_2$   & (-1.9 $\pm$ 0.7) $\times 10^{-3}$ cm$^{-1}$\\
    $A_3$   & 0.113 $\pm$ 0.02 \\
     $B_1$   & (3.3 $\pm$ 0.9) $\times 10^{-6}$ cm$^{-2}$\\
    $B_2$   & (1.4 $\pm$ 0.1) $\times 10^{-3}$ cm$^{-1}$\\
    $B_3$   & 0.984 $\pm$ 0.004 \\
    $\alpha_1$   & (1.5 $\pm$ 0.5) $\times 10^{-6}$ cm$^{-3}$\\
    $\alpha_2$   & (-5.1 $\pm$ 1.2) $\times 10^{-4}$ cm$^{-2}$\\
    $\alpha_3$   & (3.3 $\pm$ 0.8) $\times 10^{-2}$ cm$^{-1}$\\
    $\alpha_4$   & 47.3 $\pm$ 0.1 \\
     $\beta_1$   & (-1.9 $\pm$ 0.2) $\times 10^{-5}$ cm$^{-3}$\\
    $\beta_2$   & (5.3 $\pm$ 0.5) $\times 10^{-3}$ cm$^{-2}$\\
    $\beta_3$   & -0.3 $\pm$ 0.03 cm$^{-1}$\\
    $\beta_4$   & 14.1 $\pm$ 0.5 \\
          \hline
          \hline
        \end{tabular}
        \caption{Best-fit coefficients of the polynomial functions describing the $z$-dependence of the parameters in the NEST photon transit model.}
        \label{tab: photon transit coefficients value}
    \end{table}
\end{widetext}


\end{document}